\documentclass[showpacs,aps,prd,nofootinbib,floatfix,amsmath,amssymb]{revtex4-1}
\usepackage{graphicx}
\usepackage{color}
\usepackage{enumerate} 
\usepackage{multirow}

\def\beq{\begin{eqnarray}}
\def\eeq{\end{eqnarray}}

\begin{document}

\title{ Light charged Higgs boson production at the Large Hadron electron Collider }
\author{O. Flores-S\'anchez}
       \email{omar.flores@itpuebla.edu.mx}
\affiliation{Departamento de Sistemas y Computaci\'on, Tecnol\'ogico Nacional de M\'exico, 
Instituto Tecnol\'ogico de Puebla,
Av. Tecnol\'ogico num. 420, Col. Maravillas, C.P. 72220, Puebla, Puebla, M\'exico.}
\author{J. Hern\'andez-S\'anchez}
\email{jaime.hernandez@correo.buap.mx}
\affiliation{Fac. de Cs. de la Electr\'onica, Benem\'erita Universidad Aut\'onoma de Puebla, Apartado Postal 1152, 72570 Puebla, Puebla, M\'exico}
\author{C. G. Honorato }
\email{carlosg.honorato@correo.buap.mx}
\affiliation{Fac. de Cs. de la Electr\'onica, Benem\'erita Universidad Aut\'onoma de Puebla, Apartado Postal 1152, 72570 Puebla, Puebla, M\'exico}
\author{S. Moretti}
\email{s.moretti@soton.ac.uk}
\affiliation{School of Physics and Astronomy, University of Southampton, Highfield, Southampton SO17 1BJ, United Kingdom, and Particle Physics Department, Rutherford Appleton Laboratory, Chilton, Didcot, Oxon OX11 0QX, United Kingdom}
\author{S. Rosado-Navarro}
\email{sebastian.rosado@gmail.com}
\affiliation{Fac. de Cs. F\'{\i}sico-Matem\'aticas, Benem\'erita Universidad Aut\'onoma de Puebla, Apartado Postal 1364, C.P.  72570 Puebla, Puebla, M\'exico}


\date{\today}

\begin{abstract}
We  study the production of a light charged Higgs boson at the future Large Hadron electron Collider (LHeC),  through the process $e^- p \to \nu_e H^- q$,  considering both  decay channels  $H^- \to b \bar c$ and $H^- \to \tau \bar \nu_\tau$ in the final state. We analyse these processes in the context of the 2-Higgs Doublet Model Type III (2HDM-III) and assess the LHeC sensitivity to such $H^-$ signals against a variety of both reducible and irreducible backgrounds. We confirm that prospects for $H^-$ detection in the 2HDM-III are excellent assuming standard collider energy and luminosity conditions.   
\end{abstract}

\maketitle
\section{Introduction}

Now that a neutral Higgs boson has been discovered  at the Large Hadron Collider (LHC) by the ATLAS \cite{Atlas} and CMS \cite{cms} experiments, the 
SM appears to be fully established.  However, the SM-like limit of Electro-Weak Symmetry Breaking (EWSB) dynamics induced by a Higgs potential exists in several Beyond the SM (BSM) extensions of the Higgs sector. Notably,  the 
2-Higgs Doublet Model (2HDM) in its versions Type I, II, III (or Y)   and IV (or X), wherein Flavour Changing Neutral Currents (FCNCs) mediated by (pseudo)scalars can be eliminated under discrete symmetries \cite{Branco:2011iw}, is an intriguing  BSM candidate, owing to the fact that it implements the same fundamental doublet construct of the SM (albeit twice), assumes the same SM gauge group and predicts a variety of new Higgs boson states that may be accessible at the LHC. In fact, another, very interesting kind of  2HDM is  the one where FCFNs can be controlled by a particular texture in the Yukawa matrices  \cite{Fritzsch:2002ga}. In particular, in previous papers, we have implemented a four-zero-texture in a scenario which we have called 2HDM Type III (2HDM-III) \cite{DiazCruz:2009ek}. This model has a phenomenology which is very rich, which we studied at colliders in various instances \cite{Hernandez-Sanchez:2016vys}--\cite{HernandezSanchez:2013xj},
and some very interesting aspects, like flavour-violating quarks decays, which can be enhanced for  neutral Higgs bosons with intermediate mass (i.e., below  the top quark mass). {In particular, we have studied the signal $\phi_i^0 \to s \bar{b} + h.c.$ ($\phi_i^0= h, \, H$) at the future $ep$ machine LHeC \cite{Hernandez-Sanchez:2015bda, Das:2015kea}. Specifically, taking in account the characteristics of such a machine, we have established the leading charged current production process $e^-p\to \nu_e \phi_i^0 q$ followed by the signature $\phi_i^0 \to s\bar{b}+ h. c.$, by considering $3 j +  E_T \hspace{-.4 cm} / $\hspace{.2 cm}  as final state, where $j$ represents a jet and $ E_T \hspace{-.4 cm} / $\hspace{.2 cm} refers to missing transverse energy.} Furthermore, in this model, the parameter space can avoid the current experimental constraints from flavour  and Higgs physics and   a light charged Higgs  boson is allowed \cite{HernandezSanchez:2012eg}, so that  the  decay $H^- \to b \bar{c}$ is enhanced and its Branching Ratio
 ($BR$) can be dominant. In fact, this channel  has been also studied in a variety of Multi-Higgs Doublet Models 
(MHDMs) \cite{Akeroyd:2016ymd,Akeroyd:2012yg}, wherein the $BR (H^- \to b \bar{c}) \approx 0.7- 0.8$ and could afford one with a considerable gain in sensitivity to the presence of a $H^-$  by tagging the $b$ quark.  

\begin{figure} [!h]
        \centering
        \includegraphics[scale = .25]{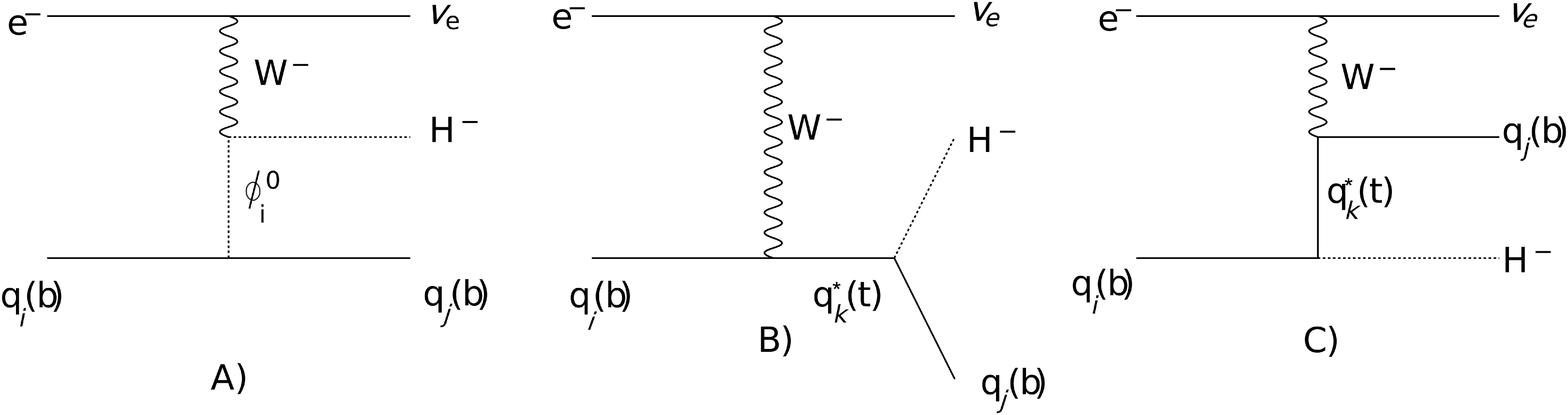}
        \caption{
		Feynman diagrams for the $e^-\ p\rightarrow \nu_e H^- q$ process.  Here, $\phi^0_i=h,H,A$, i.e., any of the neutral Higgs bosons of the BSM scenario considered here (see below). }
        \label{Feynsignal}
\end{figure}

\begin{figure} [!t]
        \centering
        \includegraphics[scale = .25]{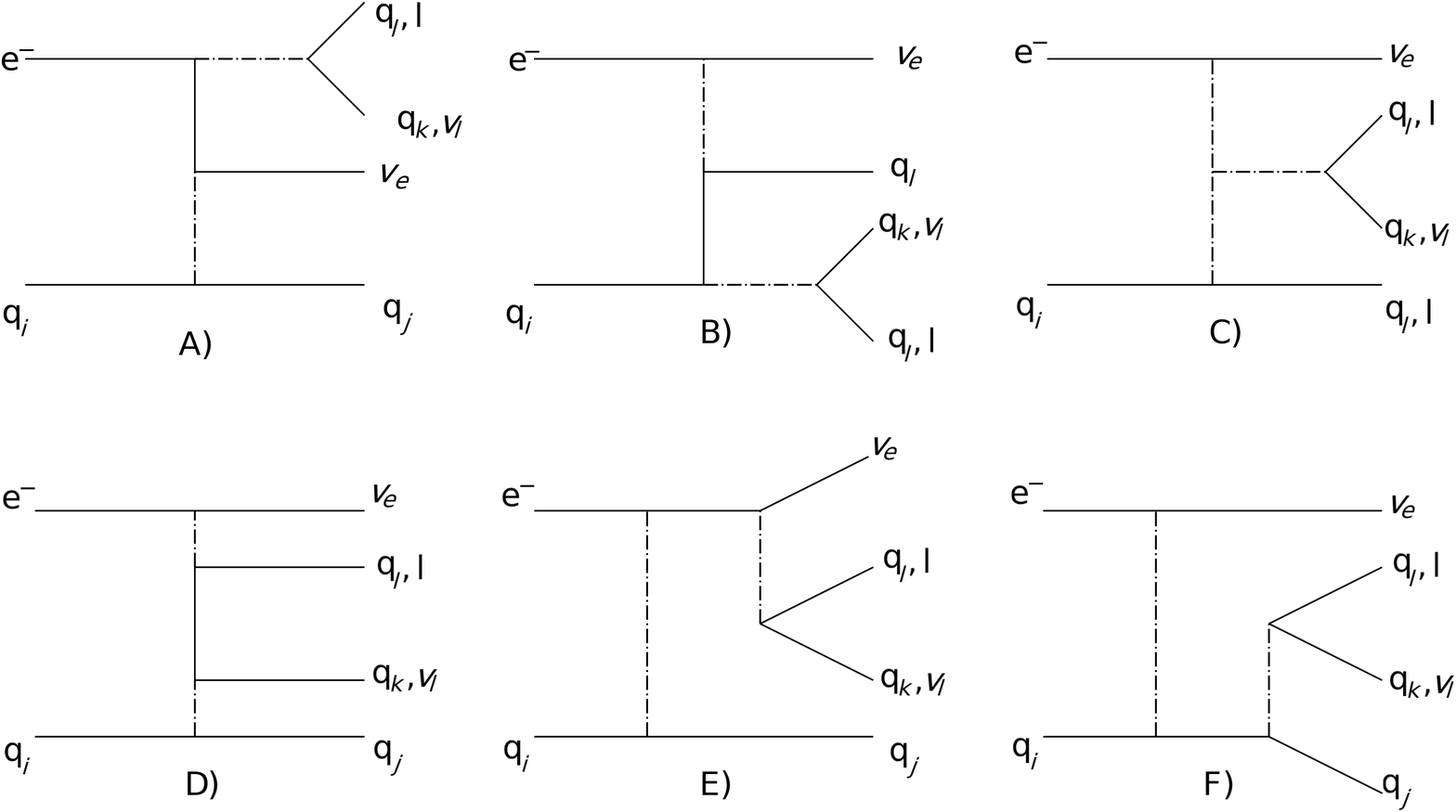}
        \caption{
		Feynman diagrams for the $\nu_e j j j$, $\nu_e b j j$ and $\nu_e b b j$ backgrounds (the change $ q_l \leftrightarrow l $ and $ q_k \leftrightarrow \nu_l $ represents the $\nu_e \nu_l l j$ and $\nu_e \nu_l l b$ backgrounds). Dash-dot lines represent boson fields:  (pseudo)scalars and EW gauge bosons.  }
        \label{FeynBG1}
\end{figure}

\begin{figure} [!b]
        \centering
        \includegraphics[scale = .25]{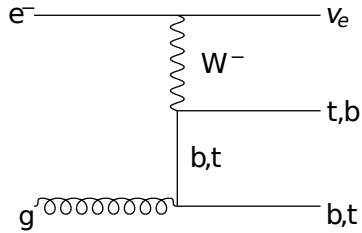}
        \caption{
		Feynman diagrams for the $\nu_e b t$ background. }
        \label{FeynBG2}
\end{figure}

{Previously, we had done a parton level study of the  process $e^-p \to \nu_e H^- b $ followed by the signal $H^- \to b\bar{c} $ \cite{Hernandez-Sanchez:2016vys}. Herein,  benchmarks scenarios had been presented, yet, they had not been subjected to the most recent experimental results from Higgs boson physics, in particular, the latest measurement of the signal strength of $ h \to b\bar{b} $ \cite{Sirunyan:2018kst}.  The complete analysis and the reconstruction of this signal at detector level for the LHeC machine is presented here.}
Furthermore, in this work, we tension the $H^- \to b \bar{c}$ channel against 
the $H^-\to \tau\bar\nu_\tau$ one and contrast the scope of the two modes in order to establish the sensitivity of the
LHeC \cite{AbelleiraFernandez:2012cc,Bruening:2013bga} to the presence of light charged Higgs bosons of the 2HDM-III. 
Specifically, we study the process $e^- p \to \nu_e H^- q$  (Fig. \ref{Feynsignal}), where $q$ represents both a light flavour $q_l=d,u,s,c$ and a $b$-quark, 
followed by the decays  $H^- \to b \bar c$ and $\tau\bar\nu_\tau$ (assuming in turn a leptonic decay of the $\tau$ into an electron or muon). In the former case, 
we compare the signal yield against that of the  main  backgrounds: $\nu 3j$, $\nu 2b j$, $\nu 2j b$  and $\nu t b$.  In the latter case, we  consider instead the  backgrounds $\nu j \ell \nu$ and $\nu b \ell \nu$. (All relevant backgrounds are schematically represented in  Figs. \ref{FeynBG1}--\ref{FeynBG2}.)

The plan of this paper is as follows. In the next section we describe the 2HDM-III. Then we discuss our results. Finally, we conclude.

\section{2HDM-III}

In the 2HDM-III, the two Higgs (pseudo)scalar doublets, $\Phi_{1}^{\dag}=( \phi_{1}^{-},\phi_{1}^{0*} ) $ and $\Phi_{2}^{\dag}=(\phi_{2}^{-},\phi_{2}^{0*})$, have hyper-charge +1 and both couple to all fermions. Here, a specific four-zero-texture is implemented as an effective flavour symmetry in the Yukawa sector,  which we have shown previously being the mechanism controlling FCNCs. Therefore, it  is not necessary to consider  discrete symmetries in the Higgs potential \cite{Felix-Beltran:2013tra,HernandezSanchez:2012eg}. Then, one can study the most general $SU(2)_L \times U(1)_Y$ invariant (pseudo)scalar potential  given by:
\begin{eqnarray}
V(\Phi_1,\Phi_2) &=& \mu_{1}^{2}(\Phi_{1}^{\dag}\Phi_{1}^{}) + \mu_{2}^{2}(\Phi_{2}^{\dag}\Phi_{2}^{}) - \left(\mu_{12}^{2}(\Phi_{1}^{\dag}\Phi_{2}^{} + h.c.)\right) \nonumber \\ 
&+& \frac{1}{2} \lambda_{1}(\Phi_{1}^{\dag}\Phi_{1}^{})^2 + \frac{1}{2} \lambda_{2}(\Phi_{2}^{\dag}\Phi_{2}^{})^2 + \lambda_{3}(\Phi_{1}^{\dag}\Phi_{1}^{})(\Phi_{2}^{\dag}\Phi_{2}^{}) + \lambda_{4}(\Phi_{1}^{\dag}\Phi_{2}^{})(\Phi_{2}^{\dag}\Phi_{1}^{}) \nonumber \\
&+& \left( \frac{1}{2} \lambda_{5}(\Phi_{1}^{\dag}\Phi_{2}^{})^2  + \lambda_{6}(\Phi_{1}^{\dag}\Phi_{1}^{})(\Phi_{1}^{\dag}\Phi_{2}^{}) + \lambda_{7}(\Phi_{2}^{\dag}\Phi_{2}^{})(\Phi_{1}^{\dag}\Phi_{2}^{}) + h.c. \right),
\end{eqnarray}
where we assume all parameters to be real\footnote{The $\mu_{12}^{2}$, $\lambda_{5}$,  $\lambda_{6}$ and  $\lambda_{7}$ parameters could be complex in general.}, including the Vacuum Expectation Values (VEVs) of the Higgs fields, hence there is no CP-Violating (CPV) dynamics. Usually, when a discrete symmetry $\Phi_1 \rightarrow \Phi_1$ and $\Phi_2 \rightarrow -\Phi_2$ is considered, the  $\lambda_6$ and $\lambda_7$ parameters are absent. In general, for two (complex)  doublet fields, there are eight fields but only five of them are physical (pseudo)scalar (``Higgs") fields, which correspond to: two neutral CP-even bosons $h$ (the light one) and $H$ (the heavy one), one neutral CP-odd boson $A$ and two charged bosons $H^\pm$. The mixing angle $\alpha$ of the two neutral CP-even bosons $h$ and $H$ is another parameter of the 2HDM. In total, the 2HDM can be described by the parameters $\alpha$, $\beta$ (where $\tan\beta$ is the ratio of the VEVs of the two doublet fields) and the masses of the five Higgs particles. With these inputs one can estimate all the parameters that are present in the scalar potential.

The other hand, when mass matrices with a four-zero-texture are considered instead, one can keep the terms proportional to $\lambda_6$ and $\lambda_7$. Besides, we have shown that these parameters play a relevant role in one-loop processes, because self-interactions of Higgs bosons are sizeable. 
In contrast, the EW parameter $\rho = m_W^2/ m_Z^2 \cos_W^2$ receives large one-loop corrections directly by the mass difference between charged Higgs and CP-even/odd masses, which can be large irrespective of the value of $\lambda_6$ and $\lambda_7$ \cite{Cordero-Cid:2013sxa}.  {Specifically, the underlying custodial symmetry(twisted custodial symmetry) is broken when the difference of the scalars masses $m_{H^\pm} -m_{A}$($m_{H^\pm} -m_{H}$) is substantial. Yet, a surviving  model is possible when the parameter $\rho \approx 1$ \cite{Gunion:2002zf,Gerard:2007kn,deVisscher:2009zb}, so we have enforced this condition.   In such a case, in agreement with Ref.  \cite{Gunion:2002zf}, when taking the SM-like scenario (when $ \cos (\beta - \alpha) \to 0$), we can obtain $m_A^2-m_H^2 = O (v^2)$ and, under these assumptions, a scenario with a light mass for the charged Higgs boson is feasible. Furthermore, the mass splitting among  $H^\pm$, $H$ and $A$ must also be reconciled with the general expressions of the oblique parameters $S,T$ and $U$ when the Higgs potential embeds CP conservation \cite{Kanemura:2011sj} (the so-called EW precisions observables \cite{pdg:2018}).   These EW bounds are implemented in the benchmarks scenarios chosen and discussed in the next section.   }

In our construction, the Yukawa Lagrangian  is given by  \cite{HernandezSanchez:2012eg}:
\begin{equation}
\label{yuklan} 
\mathcal{L}_Y = -\left( Y_{1}^{u} \bar{Q}_{L} \tilde{\Phi}_{1} u_{R} + Y_{2}^{u} \bar{Q}_{L} \tilde{\Phi}_{2} u_{R} + Y_{1}^{d} \bar{Q}_{L} \Phi_{1} d_{R} + Y_{2}^{d} \bar{Q}_{L} \Phi_{2} d_{R} + Y_{1}^{l} \bar{L}_{L} \tilde{\Phi}_{1} l_{R} + Y_{2}^{l} \bar{L}_{L} \tilde{\Phi}_{2} l_{R} \right),
\end{equation}
where  $\tilde{\Phi}_{1,2} = i\sigma_2 \Phi_{1,2}^{*}$.  So, the fermion mass matrices after EWSB are given, by: $M_f = \tfrac{1}{\sqrt{2}} \left( v_1 Y_{1}^{f} + v_2 Y_{2}^{f} \right),$ $f=u,d,l$, where we have assumed  that both Yukawa matrices $Y_{1}^{f}$ and $Y_{2}^{f}$ have the aforementioned four-zero-texture form and are Hermitian. After diagonalisation, $\bar{M}_f = V_{fL}^{\dag} M_f V_{fR}$, one has $\bar{M}_f = \tfrac{1}{\sqrt{2}} \left( v_1 \tilde{Y}_{1}^{f} + v_2 \tilde{Y}_{2}^{f} \right),$ where $\tilde{Y}_{i}^{f} = V_{fL}^{\dag} Y_{i}^{f} V_{fR}$. One can obtain a good approximation for the product $V_{q} Y_{n}^{q} V_{q}^{\dag}$ by expressing the rotated matrix $\tilde{Y}_{n}^{q}$ as  \cite{HernandezSanchez:2012eg}:
\begin{equation}
\left[ \tilde{Y}_{n}^{q} \right]_{ij} = \frac{\sqrt{m_{i}^{q} m_{j}^{q}}}{v} \left[ \tilde{\chi}_{n}^{q} \right]_{ij} = \frac{\sqrt{m_{i}^{q} m_{j}^{q}}}{v} \left[ \chi_{n}^{q} \right]_{ij} e^{i\vartheta_{ij}^{q}} ,
\end{equation}
where the $\chi$s are unknown dimensionless parameters of the model. Following  the procedure of \cite{HernandezSanchez:2012eg}, we  can  get a generic expression for the couplings of the charged Higgs bosons to the fermions  as:
\begin{eqnarray}
\mathcal{L}^{\bar{f_i}f_j\phi}= &-& \left\lbrace \frac{\sqrt{2}}{v} \bar{u}_i \left( m_{d_j} X_{ij} P_R + m_{u_i} Y_{ij} P_L \right) d_j H^{+} + \frac{\sqrt{2} m_{l_j}}{v} Z_{ij} \bar{\nu}_L l_R H^{+} + h.c. \right\rbrace, 
\end{eqnarray}
where 
$X_{ij}$, $Y_{ij}$ and $Z_{ij}$ are defined as follows:
\begin{eqnarray}
X_{ij} &=& \sum_{l=1}^{3} \left( V_{\rm CKM} \right)_{il} \left[ X \frac{m_{d_l}}{m_{d_j}} \delta_{lj} -\frac{f(X)}{\sqrt{2}} \sqrt{\frac{m_{d_l}}{m_{d_j}}} \tilde{\chi}_{lj}^{d} \right], \\
Y_{ij} &=& \sum_{l=1}^{3} \left[ Y \delta_{il} -\frac{f(Y)}{\sqrt{2}} \sqrt{\frac{m_{u_l}}{m_{u_i}}} \tilde{\chi}_{il}^{u} \right] \left( V_{\rm CKM} \right)_{lj}, \\
Z_{ij}^{l} &=& \left[ Z \frac{m_{l_i}}{m_{l_j}} \delta_{ij} -\frac{f(Z)}{\sqrt{2}} \sqrt{\frac{m_{l_i}}{m_{l_j}}} \tilde{\chi}_{ij}^{l} \right],
\end{eqnarray}
where 
$f(a)=\sqrt{1+a^2}$  and the parameters $X$, $Y$ and $Z$   are arbitrary complex numbers, which can be related to $\tan \beta$ or $\cot \beta$ when $\chi_{ij}^f=0$ \cite{HernandezSanchez:2012eg}, thus  recovering the standard four types  of the 2HDM (see the Tab.  \ref{XYZ})\footnote{So that we will refer to these 2HDM-III `incarnations' as 2HDM-III like-$\chi$ scenarios, where $\chi=$ I, II, X and Y (to be defined below).}, and the Higgs-fermion-fermion $(\phi ff)$ couplings in the 2HDM-III are written as $g_{\rm 2HDM-III}^{\phi ff} = g_{\rm 2HDM-any}^{\phi ff} + \Delta g$, where $g_{\rm 2HDM-any}^{\phi ff}$ is the coupling $\phi f f$ in any of the 2HDMs with discrete symmetry and $\Delta g$ is the contribution of the four-zero-texture. Finally, we have also  pointed out that this Lagrangian can represent a Multi-Higgs Doublet Model (MHDM) or an Aligned 2HDM (A2HDM) with additional flavour physics in the Yukawa matrices \cite{Felix-Beltran:2013tra,HernandezSanchez:2012eg}. \begin{table}[htp]
\begin{center}
\begin{tabular}{|c|c|c|c|}
\hline
2HDM-III & $X$ & $Y$& $Z$\\
\hline
2HDM Type I & $-\cot \beta$ &  $ \cot \beta$ & $- \cot \beta$ \\
\hline
2HDM Type II & $\tan \beta$ &  $ \cot \beta$ & $ \tan \beta$ \\
\hline
2HDM Type X & $-\cot \beta$ &  $ \cot \beta$ & $ \tan \beta$ \\
\hline
2HDM Type Y  & $\tan \beta$ &  $ \cot \beta$ & $ -\cot \beta$ \\
\hline
\end{tabular}
\end{center}
\caption{The parameters $X$, $Y$ and $ Z$ of the 2HDM-III defined in the Yukawa interactions when $\chi_{ij}^f=0$ so as to recover the standard four types of 2HDM.}
\label{XYZ}
\end{table}%

\section{Benchmark scenarios}
The 2HDM-III has been constrained previously by us, see Refs. \cite{Hernandez-Sanchez:2016vys}--\cite{HernandezSanchez:2013xj}, by taking into account both flavour and Higgs physics as well as  EW Precision Observables (EWPOs)
(e.g., the oblique parameters) plus  theoretical bounds such as vacuum stability, unitarity as well as  perturbativity.  {In particular,  for  our present study, we confine ourselves to the parameter space region where $m_h=125$ GeV  (hence, $h$ is the SM-like Higgs boson), with $m_A =100$ GeV,  180 GeV $< m_H < 260$ GeV and 100 GeV$<m_{H^\pm}<$ 170 GeV, further  assuming  $\cos (\beta - \alpha)=0.1, \, 0.5$.  We have fixed the oblique parameter $U=0$, since that $U$ is suppressed by a factor of order the new physics scale $\Lambda^2$ compared to the parameters $S$ and $T$ \cite{pdg:2018}, where:
$S= 0.02 \pm 0.07$ and $ T=0.06\pm 0.06$.
In general, the parameter space  of the 2HDM-III with the  four-zero-texture considered here  is fully compatible with the SM-like Higgs boson discovery \cite{Khachatryan:2016vau}, as we have implemented the same analysis of Refs. \cite{Felix-Beltran:2013tra,HernandezSanchez:2012eg,Cordero-Cid:2013sxa}, wherein the radiative decays $h\to \gamma \gamma, \, \gamma Z$ were studied and the impact of the charged Higgs bosons flowing in the corresponding loops was analysed in detail. We adapt that study here by taking into account  the most recent experimental data from the LHC for these two decay modes 
 \cite{Sirunyan:2018tbk,Sirunyan:2018koj,Aaboud:2017uhw,Aaboud:2018ezd,Aaboud:2018wps}: upon applying these filters, the mass of the light charged Higgs boson is constrained in the range  110 GeV$<m_{H^\pm}<$ 170 GeV with $\cos (\beta - \alpha)=0.1, \, 0.5$. However,  fermiophobic couplings for a charged Higgs boson with mass in the range 79 GeV$<m_{H^\pm}<$ 100 GeV are allowed in the light of the given experimental constraints. Furthermore, recent experimental bounds from flavour physics are considered here, following the analysis of Refs. \cite{Felix-Beltran:2013tra,HernandezSanchez:2012eg,Crivellin:2013wna}, where the parameter space of the model is bound by  leptonic and semi-leptonic meson decays, being the inclusive decay $B \to X_s \gamma$, $B_0-B_0$ and $K_0-K_0 $ mixing as well as $ B_s \to \mu^+ \mu^-$ transitions the strongest constraints available. One can further get a  scenario where a rather light charged Higgs mass is feasible, because the Yukawa-texture effects enter directly in the amplitudes of the mentioned mesonic decays, thus enabling one to evade these bounds. But let  us recap what are the current limits on the masses of the various  Higgs bosons from direct searches at  previous and current colliders. 
 \begin{itemize}
 \item LEP limits. A lower limit of 114 GeV was imposed for both Higgs bosons which are CP-even states, whether SM-like or not, albeit the lower mass region is ambiguous given a slight excess observed at LEP for an invariant mass around 98 GeV \cite{Barate:2003sz}.  In the MSSM configuration, for large $\tan \beta$ and low mass for the CP-odd Higgs boson, being the lightest Higgs boson $h$ non-SM-like, the limits on the neutral masses are: $m_A >93.4$ GeV and $m_h> 92.8$ GeV  \cite{Schael:2006cr}. Furthermore, from the Higgs-strahlung process, the  LEP collaborations have established  as lower bound for the heavy neutral Higgs bosons  mass of 112 GeV \cite{Kado:2002er}. For the mass of the charged Higgs boson, the LEP collaborations have instead established  a lower bound at 78.6 GeV \cite{Schael:2006cr}.
 \item Tevatron limits. The   CDF collaboration reported a local excess  in the mass region $130$ GeV $ <m_h<160$ GeV  \cite{Aaltonen:2011nh} and D0 found a local fluctuation in the $H^\pm$ mass region from 110 GeV to 125 GeV  \cite{Abazov:2010ci}. These are consistent with the later discovery of the 125 GeV state. For the case of a mass of the  charged Higgs boson in the range 90 GeV to 160 GeV, the CDF and D0 experiments extracted a limit for the BR$(t\to H^+ b)$ of $ \approx 20 \%$ considering both cases BR$(H^+\to c\bar{s})=1$ and  BR$(H^+\to \tau^+ \nu)=1$ \cite{Abbott:1999eca,Abulencia:2005jd,Abazov:2009aa}.
\item LHC limits.  The almost degenerate case for the masses of Higgs bosons in the range  110 GeV $<m_H<150$ GeV has been analysed by CMS and the experimental results can be applied in a generic way to a CP-odd state too  \cite{CMS:2013wda}. The case of additional states with exactly the same mass of the discovered Higgs boson (when it is   SM-like) is not discarded and some models  could reproduce it,  as it would be case for the 2HDM-I \cite{Heinemeyer:2013tqa}. In contrast, a fermiophobic Higgs boson with mass in the range  110 GeV to 188 GeV has been excluded by CMS    \cite{Chatrchyan:2013sfs}. Concerning CP-odd states, lately, the  CMS collaboration has reported  a slight excess with a mass just above 97.6 GeV \cite{CMS:2017yta}. However, the ATLAS experiment has not observed a corresponding significant excess \cite{pdg:2018}. Besides, recently, the CMS collaboration has excluded small values  of  $\tan \beta$ in the framework of any  2HDM in the range  225 GeV $ < m_A <$ 1000 GeV \cite{Sirunyan:2019xls}. For the analysis of a charged Higgs boson,  CMS has set  BR$(t\to H^+ b)=2-3\%$ as upper limit for the case BR$(H^+\to \tau^+ \nu)=1$ in the range of masses varying from  80 GeV to 160 GeV \cite{pdg:2018}. Otherwise, assuming   BR$(H^+\to c\bar{s})=1$, ATLAS and CMS set BR$(t\to H^+ b)\approx 20 \%$ for the mass range 90 GeV to 160 GeV \cite{pdg:2018}. Finally, quite recently, CMS set a limit of  BR$(t\to H^+ b)=0.5-0.8\%$ for the case BR$(H^+\to c\bar{b})=1$ in the mass range 90 GeV to 150 GeV \cite{Sirunyan:2018dvm}.
 \end{itemize}

 So, considering all experimental bounds and theoretical constraints, we proceed to choose several  scenarios.}   
Specifically, we consider four scenarios, wherein relevant Benchmarks Points (BPs) are defined  according to the standard Yukawa prescriptions:  Type I (where one Higgs doublet couples to all fermions); Type II (where one Higgs doublet couples to the up-type quarks and the other to the down-type quarks);  Type X (also called IV or "Lepton-specific", where the quark couplings are Type I and the lepton ones are Type II); Type Y (also called III or "Flipped" model, where the quark couplings are Type II and the lepton ones are Type I). 

For a light charged Higgs boson, in the 2HDM-III,  the most important decay channels are $H^- \to s\bar c$ and $ b \bar c$, when $Y \gg X, Z$ (like-I scenario), $X, \, Z \gg Y$ (like-II scenario) or $X \gg Y, \, Z$  (like-Y scenario), in which cases the mode $H^-\to b\bar c$ receives  a substantial enhancement, coming from the 
four-zero-texture implemented in the Yukawa matrices, so as to obtain even a BR$(H^- \to b\bar c) \approx 0.95$, so that we focus on this decay, also owing to the fact that it can be $b$-tagged, thus reducing in turn the level of background. However, for the case $Z \gg X, Y$ (like-X scenario),  the  decay channel $H^-  \to \tau \bar\nu_\tau$ is maximised, reaching a $B$R of 90 \%  or so \cite{HernandezSanchez:2012eg}, so that we will investigate this mode as well.  

In this work,  considering the parameter scan performed in \cite{Das:2015kea}, we adopt the following BPs, where the aforementioned two decay channels
($H^- \to b\bar c$ and $H^-  \to \tau \bar\nu_\tau$) offer the most optimistic chances for detection.
\begin{itemize}
\item Scenario 2HDM-III like-I: $\cos (\beta- \alpha)= 0.5$, $\chi^u_{22}= 1$, $\chi^u_{23}= 0.1$, $\chi^u_{33}= 1.4$, $\chi^d_{22}= 1.8$, $\chi^d_{23}= 0.1$,
$\chi^d_{33}= 1.2$,  $\chi^\ell_{22} =-0.4, \chi^\ell_{23}= 0.1$, $\chi^\ell_{33} =1$ with $Y \gg X, \, Z$.
\item Scenario 2HDM-III like-II: $\cos (\beta- \alpha)= 0.1$, $\chi^u_{22}= 1$, $\chi^u_{23}= -0.53$, $\chi^u_{33}= 1.4$, $\chi^d_{22}= 1.8$, $\chi^d_{23}= 0.2$,
$\chi^d_{33}= 1.3$,  $\chi^\ell_{22} =-0.4, \chi^\ell_{23}= 0.1$, $\chi^\ell_{33} =1$ with $X, \, Z \gg Y$.
\item Scenario 2HDM-III like-X: the same parameters of scenario 2HDM-III like-II but  $Z \gg X, \, Y$.
\item Scenario 2HDM-III like-Y: the same parameters of scenario 2HDM-III like-II but  $X \gg Y, \, Z$.
\end{itemize}
For all four benchmarks scenarios, we assume $m_h=125$ GeV and consider $m_A =100$ GeV,  $ m_H = 190$ GeV and 100 GeV$<m_{H^\pm}<$ 170 GeV. 

\begin{figure} [!t]
        \centering
        \includegraphics[scale = .5]{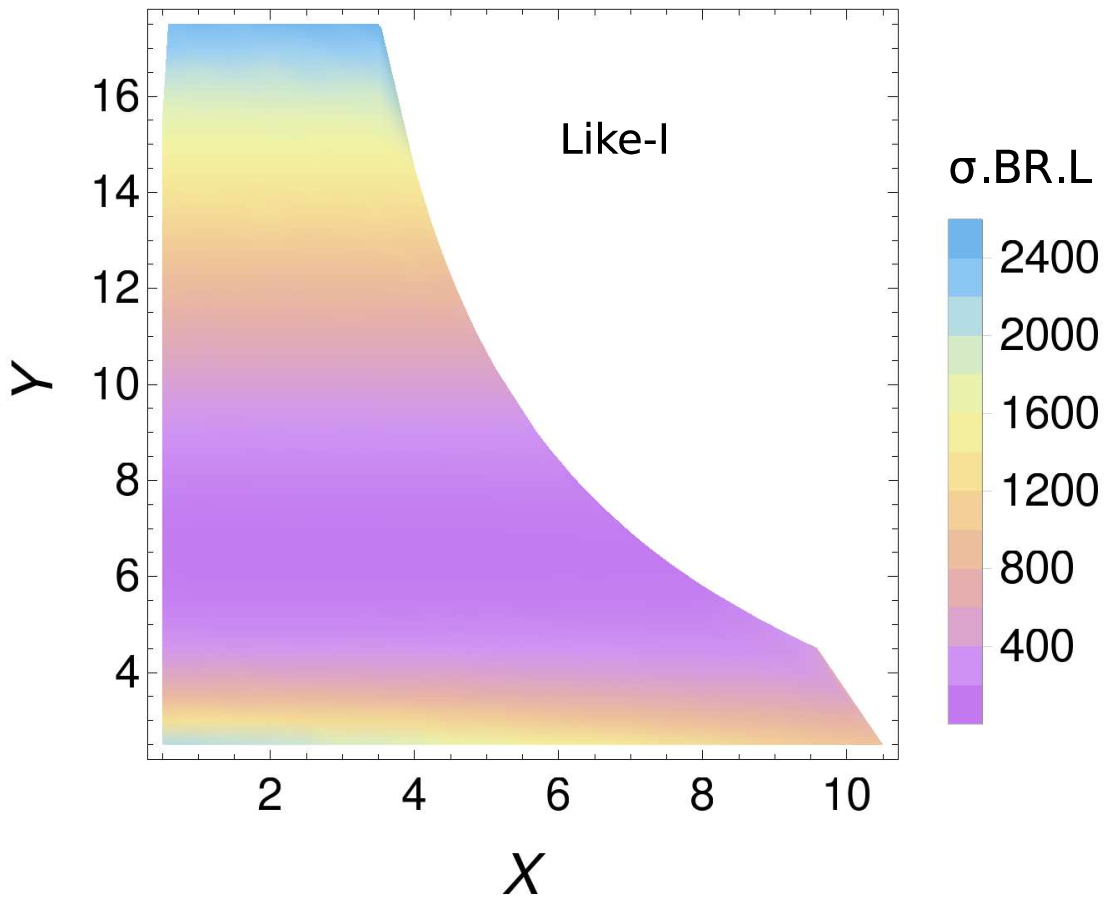}
         \includegraphics[scale = .5]{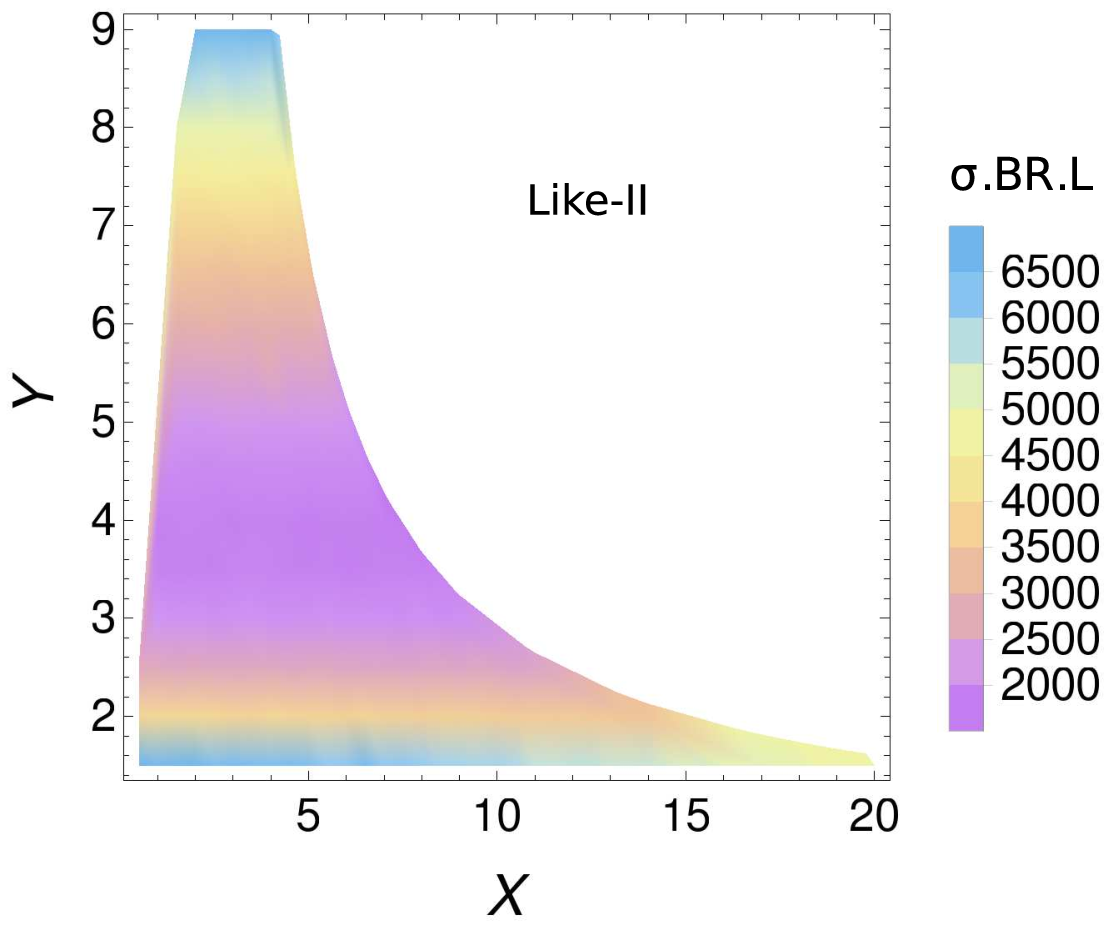}
         \includegraphics[scale = .5]{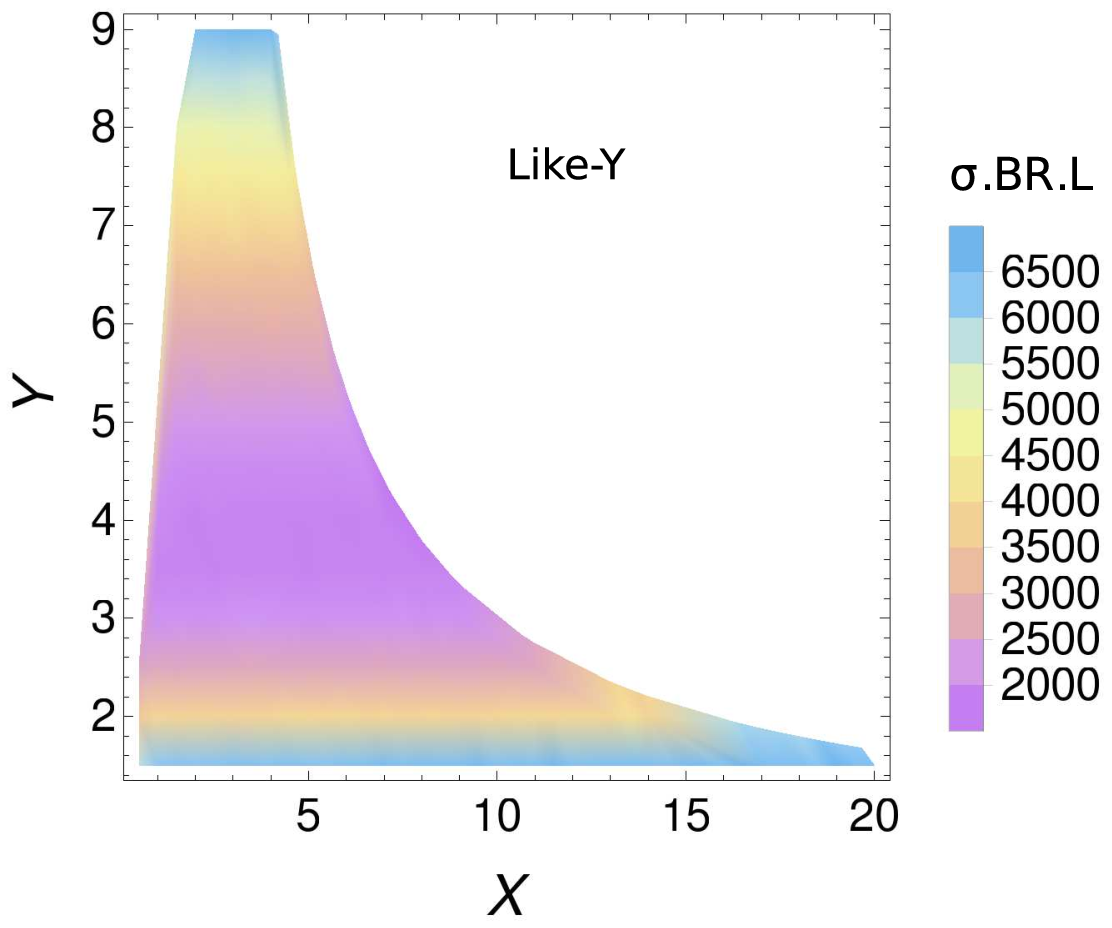}
        \caption{
		Event rates $\sigma . BR.L$ at the LHeC with $\sqrt{ s_{ep}} \approx 1.3$ TeV, where $\sigma\equiv\sigma(ep\to \nu_e H^- q)$ with $q=q_l$ or $b$ is the production cross section, 
$L=100$ fb$^{-1}$ is the integrated luminosity  and BR is the  decay fraction for the channel $H^- \to b \bar{c} $, for the following 2HDM-III scenarios: 
like-I (left),  like-II (centre) and like-Y (right).  
		}
        \label{SigmaBRScan-cb}
\end{figure}

\begin{figure} [!h]
        \centering
                 \includegraphics[scale = .6]{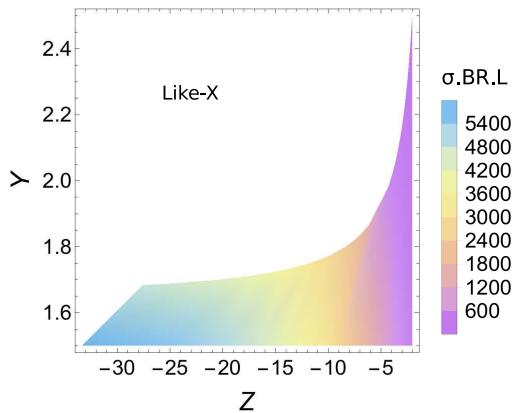}
        \caption{
		Event rates $\sigma . BR.L$ at the LHeC with $\sqrt{ s_{ep}} \approx 1.3$ TeV, where $\sigma\equiv\sigma(ep\to \nu_e H^- q)$ with $q=q_l$ or $b$ is the production cross section, 
$L=100$ fb$^{-1}$ is the integrated luminosity  and BR is the  decay fraction for the channel $H^- \to \tau \bar{\nu}_\tau $, for the following 2HDM-III scenario: like-X.}
        \label{SigmaBRScan-taunu}
\end{figure}

Before proceeding to investigate the aforementioned two $H^-$ decays, in order to gain some insights into the inclusive event rates available, we show in Figs. \ref{SigmaBRScan-cb} and \ref{SigmaBRScan-taunu}  a scan over the relevant parameters $X, Y$ and $Z$ for the four 2HDM-III incarnations, each in correspondence of the relevant $H^- \to b \bar{c}$ and $H^- \to \tau\bar\nu_\tau$ decay channels, respectively. Assuming the LHeC standard Centre-of-Mass (CM) energy of $\sqrt{ s_{ep}} \approx 1.3$ TeV and luminosity of $L = 100$ fb$^{-1}$, it is clear that inclusive event rates are substantial, of order up to several thousands in all four cases, so that the potential of the LHeC in extracting the $H^- \to b \bar{c}  $ and  $H^- \to \tau \nu_\tau $ decays is definitely worth exploring further. In fact, the main objective of our analysis is to tension one decay against the other and extract the corresponding significances, which may lead to evidencing or indeed discovering the true underlying 2HDM structure onsetting EWSB.


\section{Discussion}

As intimated, in the framework of the  2HDM-III considered here, there are two main $H^\pm$  decay channels, which are
$H^-\rightarrow b\bar c$ (the leading one for the incarnations like-I, -II and -Y) and
$H^-\rightarrow \tau \bar\nu_{\tau}$ (the leading one for the incarnation like-X). Some BPs, maximising the signal rates in the four 2HDM-III incarnations defined in terms of the parameters $\chi_{ij}^f$  and $X,Y$ and $Z$ introduced previously, are given in Tab. \ref{tab:BR}, wherein the relevant $BR$s of the $H^\pm$ state are given alongside the cross sections of the associated production process $ep\to \nu_e H^- q$, where $q=q_l$ or $b$. (However, we have eventually verified that only the case $q=b$ is phenomenologically relevant, so that, henceforth, we neglect discussing the case $q=q_l$ explicitly, though it is included in our simulations.)

The signatures that we will consider are as follows.

\begin{itemize}
\item On the one hand, in connection with the 2HDM-III  like-I, -II and -Y, wherein the most relevant decay process is {$H^-\to b\bar{c}$}, the  final state is  $3j+E_T \hspace{-.4 cm} / $\hspace{.2 cm} (where $j$ is a generic jet and $E_T \hspace{-.4 cm} / $\hspace{.2 cm} refers to missing transverse energy), with  one $b$-tagged and one light jet  (associated to the  charged Higgs boson reconstruction) accompanied by a remaining jet which can be $b$-tagged or not. 

 \item On the other hand, in connection with the 2HDM-III  like-X, wherein the most relevant decays process is {$H^-\to \tau\bar\nu_\tau$}, the final state is   $j+l+E_T \hspace{-.4 cm} / $\hspace{.2 cm}, where $l=e,\mu$ (from a leptonic $\tau$ decay) and the jet is $b$-tagged.
\end{itemize}

In this upcoming discussion we will describe the phenomenology of these two possible processes. In order to carry out our numerical analysis, we have used CalcHEP 3.7 \cite{Belyaev:2012qa}  as Parton level
 event generator, interfaced to the CTEQ6L1 Parton Distribution Functions (PDFs) \cite{Pumplin:2002vw}, 
then PYTHIA6  \cite{Sjostrand:2006za} for the Parton shower, hadronisation and hadron decays and 
 PGS \cite{PGS} as detector emulator, by using  a  LHC parameter 
card suitably modified for the LHeC \cite{AbelleiraFernandez:2012cc,Bruening:2013bga}. In particular, the detector parameters simulated were as follows:  we  considered a calorimeter 
coverage $|\eta|<5.0$, with segmentation $\Delta\eta\times\Delta\phi=0.0359\times 0314$ (the number of division in $\eta$ and $\phi$ are 320 and 200, respectively). Moreover, we used Gaussian energy resolution, with
\begin{equation}
\frac{\Delta E}{E}=\frac{a}{\sqrt{E}}\oplus b,
\end{equation}
 {where $a=0.085$ and $b=0.003$ for the Electro-Magnetic (EM) calorimeter resolution and $a=0.32$, $b=0.086$ for the hadronic calorimeter resolution, with $\oplus $ meaning addition in  quadrature. Herein, the values of $a$ and $b$  are parameters established by the design of the LHeC \cite{AbelleiraFernandez:2012cc,Bruening:2013bga}}. The algorithm to perform jet finding was a``cone" one with jet radius $\Delta R=0.5$. The calorimeter trigger cluster finding a seed(shoulder) threshold was $5$ GeV($1$ GeV). We  took $E_T(j)>10 $ GeV for a jet  to be considered so, in addition to the isolation criterion $\Delta R(j;l)>0.5$. Finally,   we have mapped the kinematic behaviour of the final state particles using  MadAnalysis5 \cite{Conte:2012fm}.

\begin{table}[!t]
	\begin{center}
		\begin{tabular}{|c|c|c|c|c|c|c|c|c|c|c|c|}
		\hline
		2HDM-III & \multicolumn{3}{|c|}{Parameters} & \multicolumn{4}{|c|}{$\sigma(ep\to\nu_eH^- q)$ (pb)}
		 & BR$(H^-\to b\bar c)$ &  BR$(H^-\to \tau \bar\nu_\tau)$\\
		\cline{2-8}
		like- & ${X}$ & ${Y}$ & ${Z}$ &{$m_{H^\pm}=110$ GeV}&{$130$ GeV}&
		{$150$ GeV} &{$170$ GeV}&{$m_{H^\pm}=110$ GeV}
		&{$m_{H^\pm}=110$ GeV}\\
		\hline
		\hline
		I & {0.5} & {17.5} & {0.5}&{$2.56\times 10^{-2}$} &{$1.30\times 10^{-2}$}
		&{$3.47\times 10^{-3}$}&{$1.35\times 10^{-4}$}&{$9.57\times 10^{-1}$}
		&{$2.5\times 10^{-4}$}\\
		\hline
		II & {20}& {1.5} & {20}& {$2.18\times 10^{-2}$}& {$1.13\times 10^{-2}$}
		& {$2.95\times 10^{-3}$} & {$5.89\times 10^{-5}$} & {$9.9\times 10^{-1}$}& 
		{$2.22\times 10^{-4}$}\\
		\hline
		X & {0.03}& {1.5} & {$-33.33$}& {$6.49\times 10^{-2}$}&{$3.39\times 10^{-2}$}
		& {$8.83\times 10^{-3}$} & {$2.34\times 10^{-4}$} & {$9.28\times 10^{-2}$}& 
		{$9.04\times 10^{-1}$}\\
		\hline
		Y & {13}& {1.5} & {$-1/13$}& {$6.41\times 10^{-2}$}& {$3.27\times 10^{-2}$}
		& {$8.47\times 10^{-3}$} & {$2.2\times 10^{-4}$} & {$9.91\times 10^{-1}$}& 
		{$6.12\times 10^{-3}$}\\
		\hline
		\end{tabular}
		\caption{The BPs that we studied for the 2HDM-III  in  the incarnations like-I, -II, -X and -Y. We present cross sections and $BRs$ at Parton level, for some $H^\pm$ mass choices. 
		}
		\label{tab:BR}
	\end{center}
\end{table}

\subsection{The  process  $e^-q\to \nu_e H^- b$  with {$H^-\to b \bar{c}$} for the 2HDM-III  like-I, -II and -Y}

In this subsection we discuss the  final state with  one $b$-tagged jet and one light jet (associated with the secondary decay $H^-\to b \bar c $) alongside a generic (i.e., light or $b$-tagged)
forward jet (associated with the primary collision) plus missing transverse energy. For this case, we apply the following cuts\footnote{For illustration, we assume the 2HDM-III like-Y scenario in our description, though the signal kinematics is essentially independent of the theoretical setup, as it primarily depends on the $m_{H^\pm}$ value.}.

\begin{figure} [t]
        \centering
        \includegraphics[scale=.45]{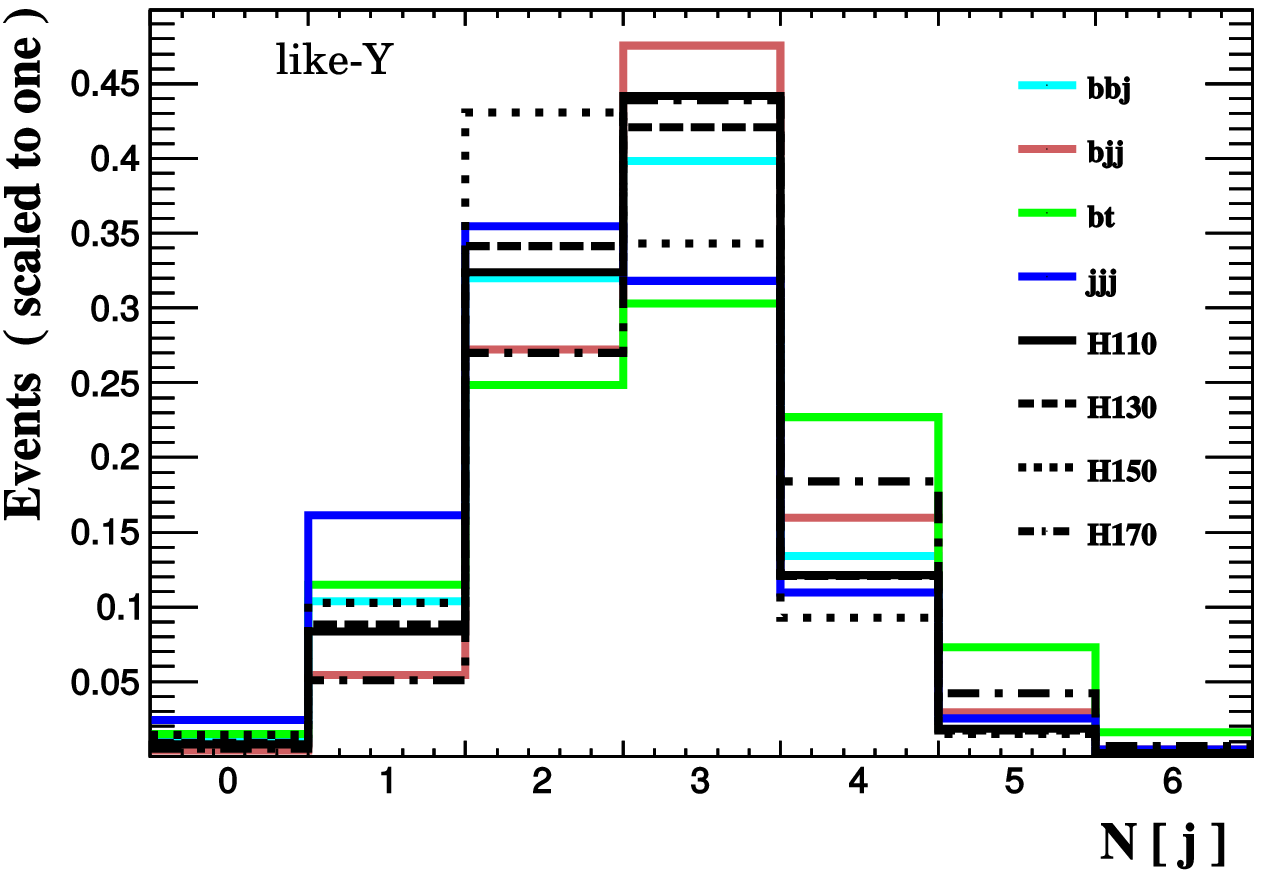}
	\includegraphics[scale=0.45]{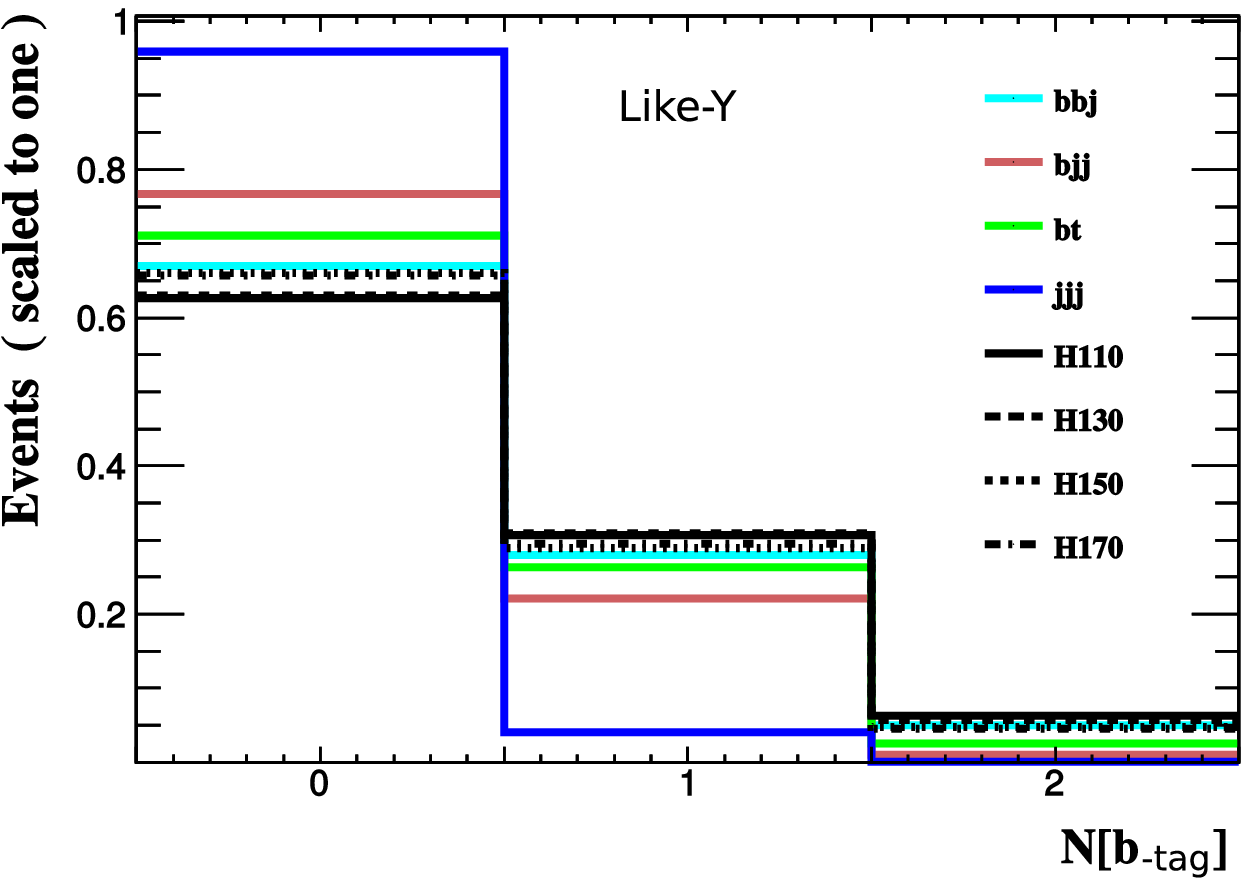}
        \caption{Distributions for the  process  $e^-q\to \nu_e H^- b$  followed by {$H^-\to b \bar{c}$}: in the  left panel we present the multiplicity of all jets while in the right panel we present the multiplicity of the  $b$-tagged ones. The like-Y case is illustrated. The normalisation is to unity.}
        \label{his-N}
\end{figure}

\begin{figure} [t]
        \centering
        \includegraphics[scale=.45]{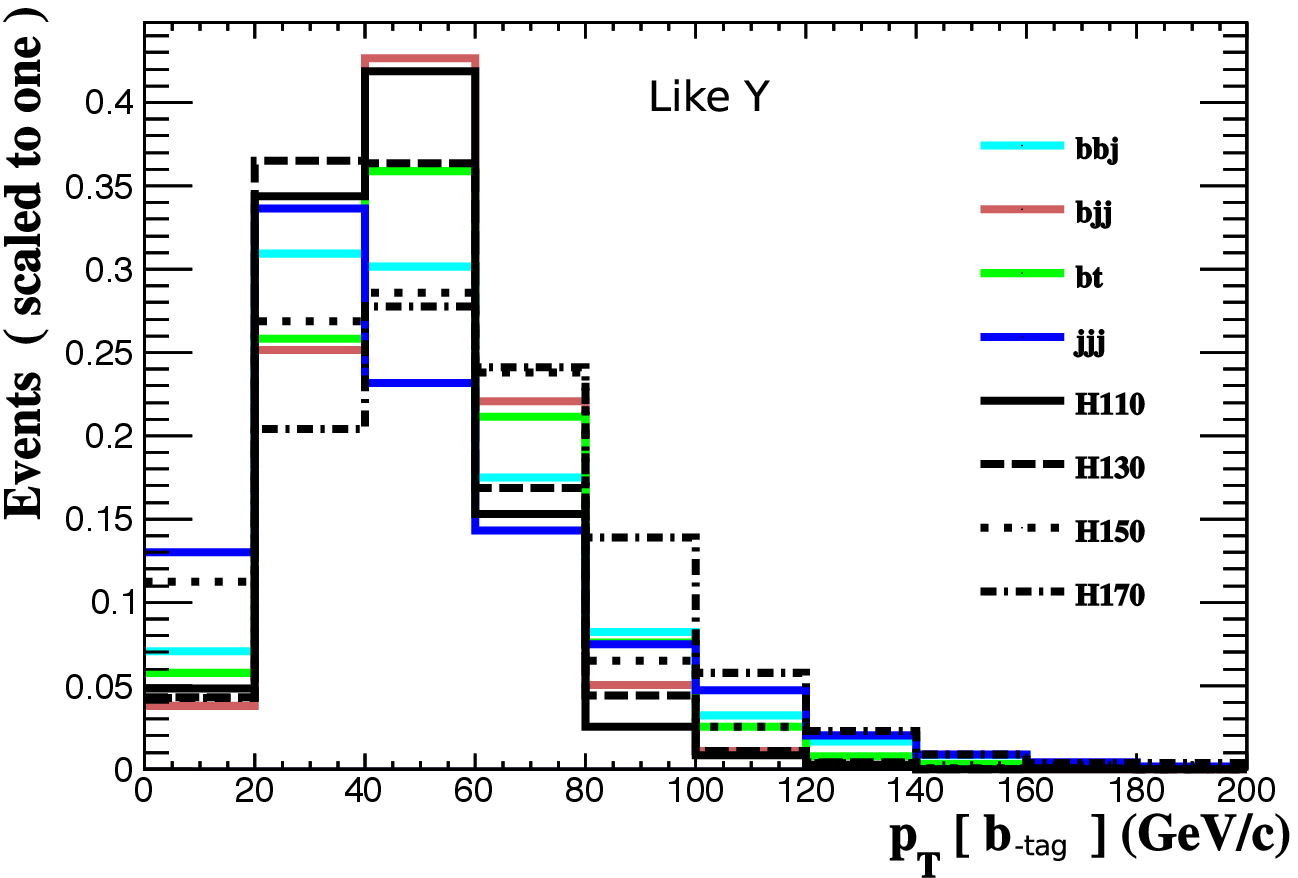}
	\includegraphics[scale=0.45]{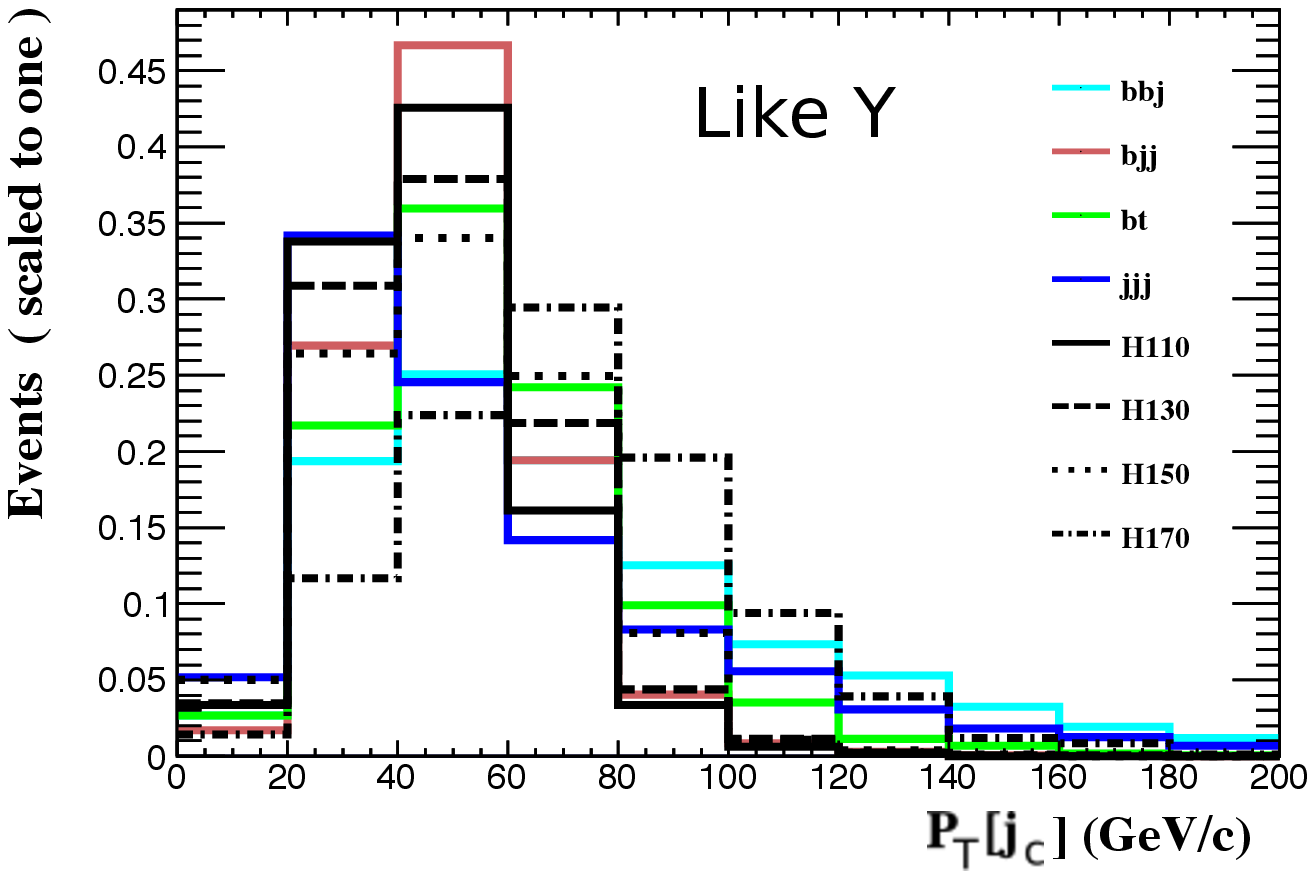}\\
	\includegraphics[scale=0.45]{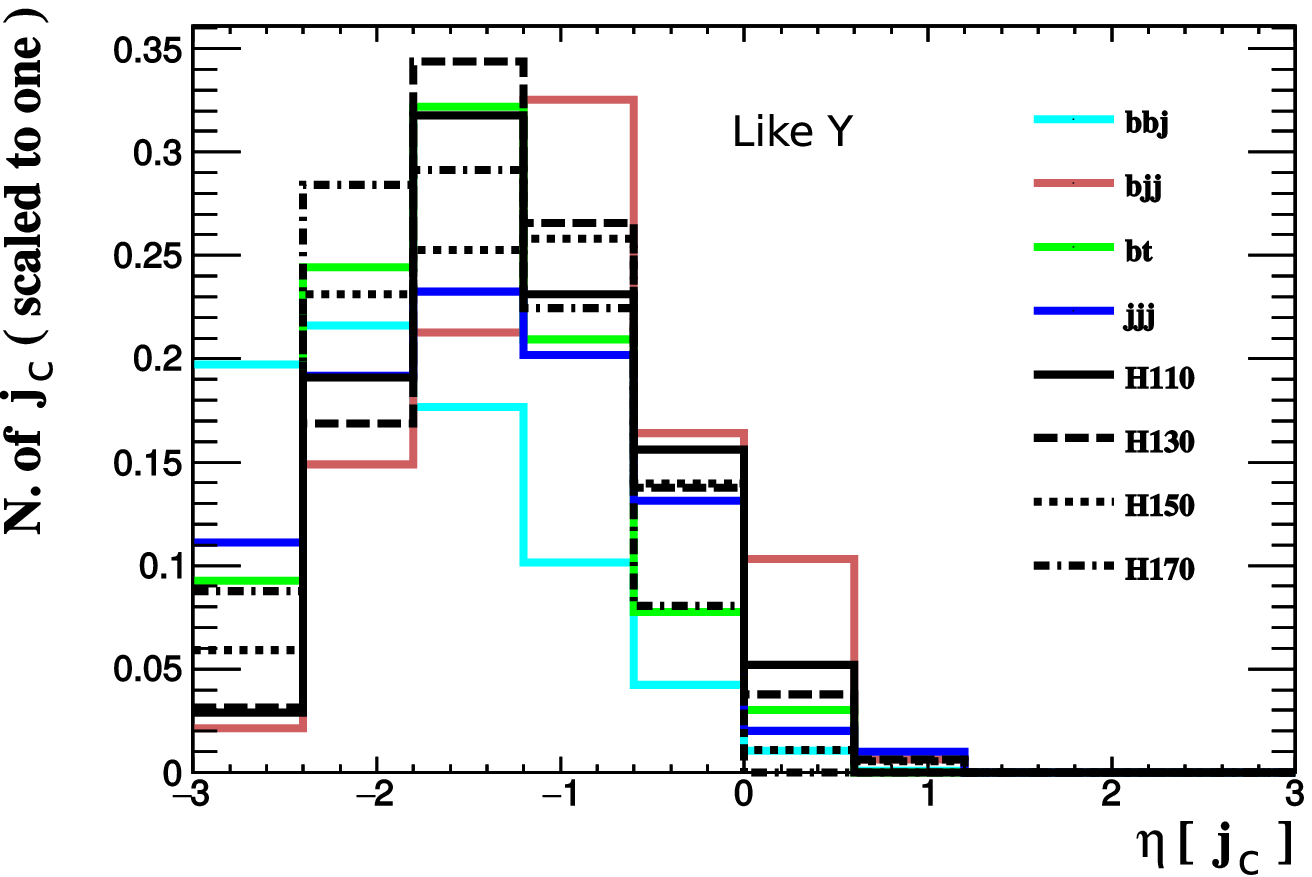}
	\includegraphics[scale=0.45]{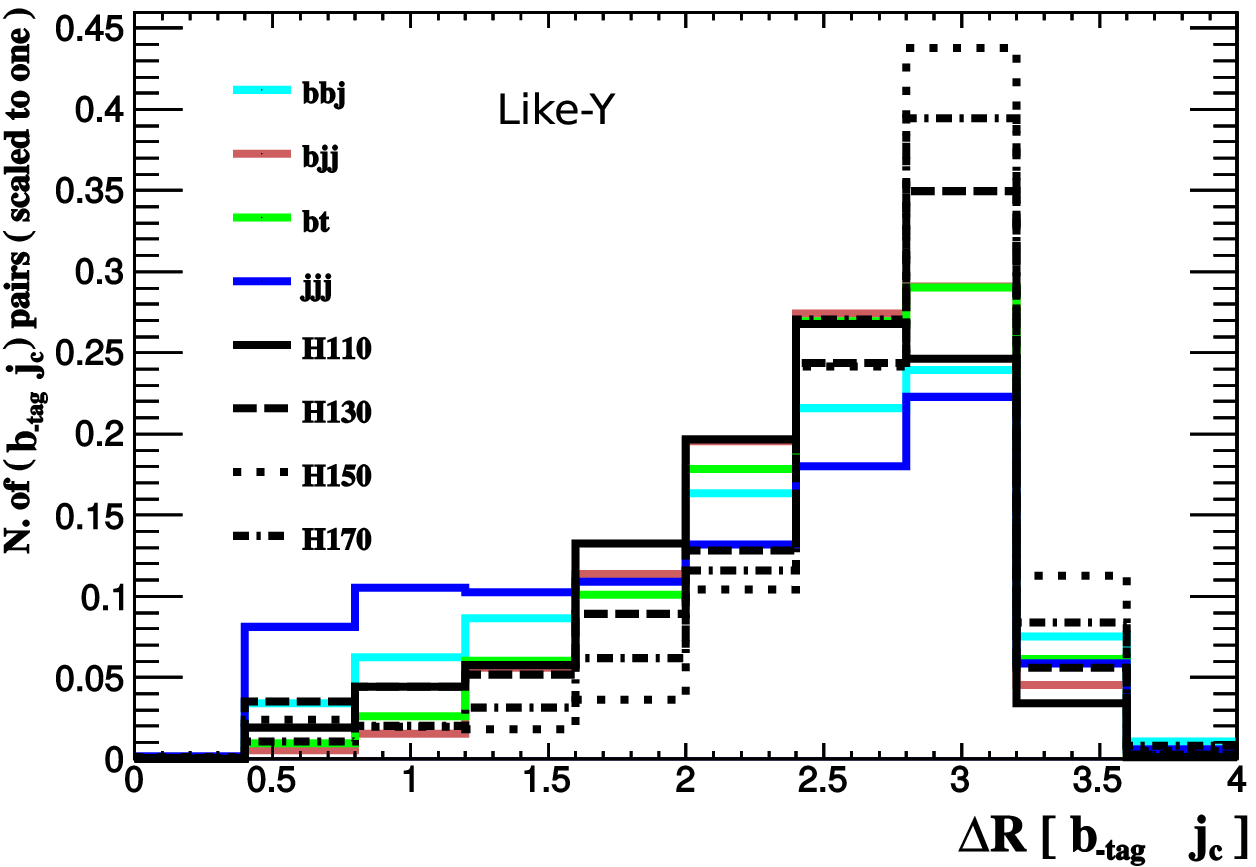}
        \caption{Distributions for the  process  $e^-q\to \nu_e H^- b$  followed by {$H^-\to b \bar c$}:
in the top-left panel we present the transverse momentum of the central  $b$-tagged jet, in the top-right panel we present the transverse momentum of the central light jet, in the bottom-left panel we present the pseudorapidity of the central light jet while in the bottom-right panel we present the separation between the two central jets. 
The like-Y case is illustrated. The normalisation is to unity.}
        \label{his-bb1}
\end{figure}

\begin{figure}
        \centering
        \includegraphics[scale=.45]{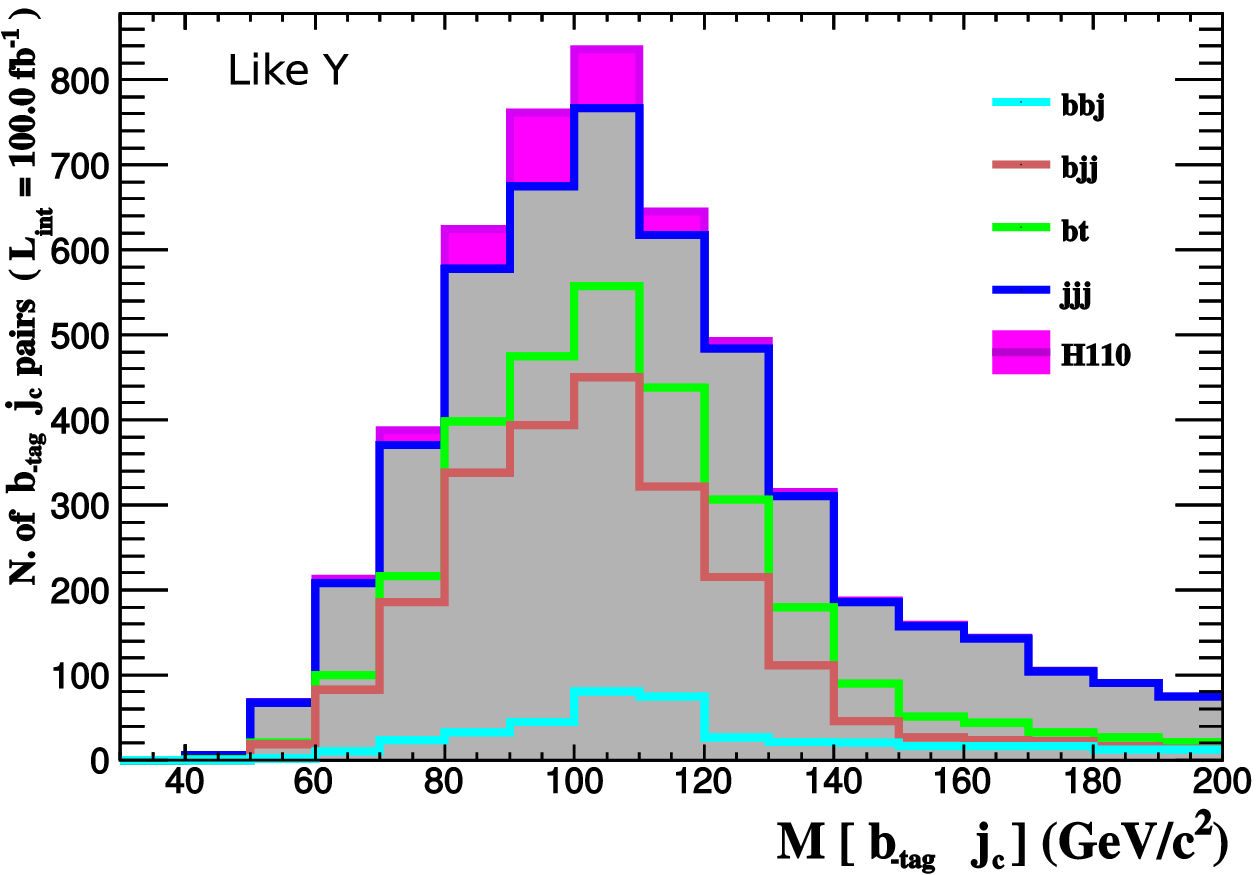}
        \includegraphics[scale=.45]{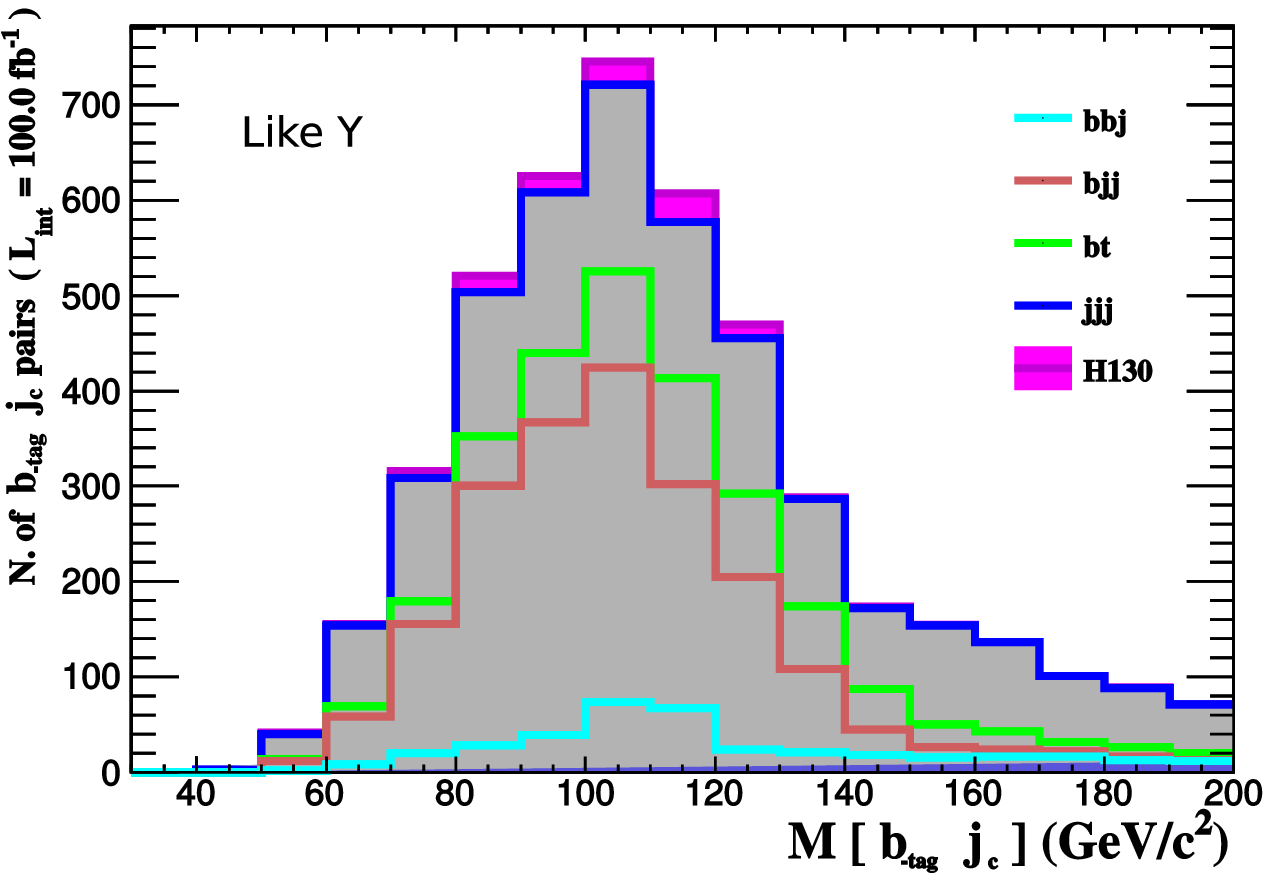}
        \caption{Distributions for the  process  $e^-q\to \nu_e H^- b$  followed by {$H^-\to b\bar c$} in the invariant mass of the two central  jets for $m_{H^\pm}=110$ GeV (left) and  $m_{H^\pm}=130$ GeV (right). 
The like-Y case is illustrated. The normalisation is to the total event rate for $L=100$ fb$^{-1}$.}
        \label{his-M2}
\end{figure}

\begin{enumerate} [I)]
\item First, we select only events with exactly three jets in the final state. Then, we reject all events without a  $b$-tagged jet. Hence, at this point, we 
keep events like $3j+E_T \hspace{-.4 cm} / $\hspace{.2 cm} with at least one $b$-tagged jet (see the histograms in 
Fig. \ref{his-N}). 
For these selections, our signal generally has an efficiency of $12\%$ while the most efficient  background  
$\nu_ebbj$ has a $10\%$ response. The remaining backgrounds have efficiencies of $5\%$, $8\%$ and $1\%$ for $\nu_e 
bt$, $\nu_e bjj$  and $\nu_e jjj$, respectively. 

\item The second set of cuts is focused on selecting two jets (one $b$-tagged, labelled as $b_{\rm tag}$, and one not, labelled as $j_{\rm c}$) which are central in the detector. First, we  demand that $P_T(b_{\rm tag})>30(40)$ GeV and $P_T(j_{\rm c})>20(30)$ GeV for $m_{H^\pm}=110,130(150,170)$ GeV (here, $P_T$ is the transverse momentum).   Then, we impose a cut on the pseudorapidity $|\eta(b_{\rm tag},j_{\rm c})|<2.5$ of both these jets and, finally, select events in which $1.8(2)<\Delta R(j_{\rm c};b_{\rm tag})<3.4(3.4)$ in correspondence of $m_{H^\pm}=110,130(150,170)$ GeV (where $\Delta R$ is the standard cone separation). Upon enforcing these cuts, we find that our signal has a cumulative efficiency of $7.3\%$. The most efficient background $\nu_ebjj$ has a rate of $6\%$ while the others show efficiencies of  $3.3\%$,  $3.7\%$ and $0.3\%$ (for $\nu_e bt$, $\nu_e bb j$ and $\nu_e jjj$, respectively). This information is easily drawn from Fig. \ref{his-bb1}.  

\item The next cut is related to the selection of a forward third generic jet (it can be either  a light jet or a $b$-tagged one). 
 Our selection for such a third jet is   $|\eta|>0.6 $ (with a transverse momentum above 20 GeV). With this cut,  our signal shows an efficiency of $5.4\%$ while $4.2\%$ is the rate for the most efficient background ($\nu_e bjj$). The rest of the backgrounds show efficiencies below  $2\%$ for $
 \nu_e bbj$ and $\nu_e bt$ or $0.3\%$ for $\nu_e jjj$.

\item The selection of the  jet pair representing a $H^{\pm}$ candidate is made by considering only events for which the invariant mass of the two central jets is in the vicinity of the (trial) mass of the charged Higgs boson. 
However, it must be considered that, at the detector level, the signal may see a mass shift due to the finite efficiency in selecting the wanted jet dynamics. 
Therefore, in the histograms of Fig. \ref{his-M2}, we study such invariant mass in the case of our signal for, e.g., $m_{H^\pm}=110$ (left) and 130 (right) GeV. We benchmark these against the corresponding spectra from the backgrounds.  From this plot, we can indeed see a shift of the signal peaks towards lower invariant masses, so that we can implement the following selection criterium: 
 $m_{H^\pm}-20$ GeV $<M(b_{\rm tag},j_{\rm c})< m_{H^\pm}$. Furthermore, we noticed that  the invariant mass formed by the  
 light central jet and the generic forward jet (not shown here) has a structure in most of the backgrounds, dictated by the presence of a hadronic $W^{\pm}$ boson decay. Because our signal does not have this feature,  we further impose that 
 $M (j_{\rm c},j_{\rm f})>80$ GeV or $M(j_{\rm c},j_{\rm f})<60$ GeV (where $j_{\rm f}$ labels the forward jet). This combination of mass cuts is highly selective, giving us an overall efficiency of $2.4\%$ for the signal and (at most) $0.6\%$  for the backgrounds. 
\end{enumerate}

\begin{table}
	\begin{center} 
	\label{cuts1}
		\begin{tabular}{|r|c|c|r|r|r|r|c|}
		\hline
		Signal & Scenario & Events (raw) & Cut I & Cut II & Cut III &Cut IV  & ${\cal(S/\sqrt B)}
		_{100\,{\rm fb}^{-1} (1000\, {\rm fb}^{-1}) [3000\, {\rm fb}^{-1}]} $ \\
		\hline
		\hline
		$\nu_e H^{\pm} b$&I-110 & 2562 & 298 & 182 & 134 & 54  & 1.43 (4.52) [7.82]\\
				&I-130 & 1300 & 139 & 82 & 64 & 19  & 0.58 (1.82) [3.16]\\
				&I-150 & 347 & 29 & 13 & 11 & 3 & 0.16 (0.5) [0.86]\\
				&I-170 & 13 & 1.29 & 0.62 & 0.51 & 0.14  & 0.01 (0.03) [0.05]\\
                     \hline
		$\nu_e H^{\pm} b$&II-110 & 2183 & 245 & 151 & 122 & 53  & 1.4 (4.43) [7.68]\\
				&II-130 & 1128 & 128 & 84 & 71 & 22  & 0.7 (2.21) [3.82]\\
				&II-150 & 294 & 28 & 14 & 13 & 4  & 0.2 (0.65) [1.13]\\
				&II-170 & 6 & 0.6 & 0.33 & 0.3 & 0.08  & 0.005 (0.017) [0.029]\\
		\hline
		$\nu_e H^{\pm} b$&Y-110 & 6417 & 468 & 567 & 347 & 156 & 4.18 (12.99) [22.5]\\
				&Y-130 & 3268 & 366 & 204 & 156 & 46 & 1.43 (4.53) [7.84]\\
				&Y-150 & 847 & 68 & 29 & 23 & 6  & 0.33 (1.06) [1.83]\\
				&Y-170 & 22 & 2.3 & 1.12& 0.89 & 0.25  & 0.017 (0.05) [0.09]\\
		\hline
		$\nu_e bb j$& & 20169 & 2011 & 748 & 569 & 125&\\
		\cline{1-7} 
		$\nu_e b jj$& & 117560 & 10278 & 7211 & 5011 & 718 & ${\cal B}=1441$\\
		\cline{1-7}
		$\nu_e b t$&  & 41885 & 2278 & 1418 & 1130 & 188 & $\sqrt{{\cal B}}=37.9$\\
		\cline{1-7} 
		$\nu_e jjj$&  & 867000 & 9238 & 3221& 2593 &  409 &\\ 
		\hline
		\end{tabular}
\caption{Significances obtained  after the sequential cuts described in the text for the signal process  $e^-q\to \nu_e H^- b$  followed by {$H^-\to b \bar{c}$} for four BPs in the  2HDM-III  like-I, -II and -Y. The simulation is done at detector level. {In the column Scenario,  the label A-110(130)[150]\{170\}  means  $m_{H^\pm} =110$(130)[150]\{170\} GeV in the 2HDM-III like-A, where A can be I, II and Y.} }
		\label{cb2}
	\end{center}
\end{table}

The final results, following the application of Cuts I--IV, are found in Tab. \ref{cb2}, for the 2HDM-III like-I, -II and -Y incarnations. Statistically, significances of the signal ${\cal S}$ over the cumulative background ${\cal B}$ are very good at low $H^\pm$ masses already for 100 fb$^{-1}$ of luminosity. As the latter increases, larger masses can be afforded through  evidence or discovery, particularly so in the like-Y scenario. However, an ultimate mass reach is probably 130 GeV in all cases.

\subsection{The  process  $e^-q\to \nu_e H^- b$  with $H^-\rightarrow \tau \bar \nu_\tau$ in the 2HDM-III like-X}

Now we focus our attention on the channel $H^- \to \tau\bar\nu_\tau$. To this effect, as previously mentioned, we look at leptonic $\tau$ decays ($\tau\to l\bar\nu_l \nu_\tau$, with $l=e,\mu$) and we  $b$-tag the prompt (i.e., coming from the primary collision) jet in the final state.
The  cuts to extract  our signal are presented below.

\begin{figure}[!t]
        \centering
        \includegraphics[scale=.45]{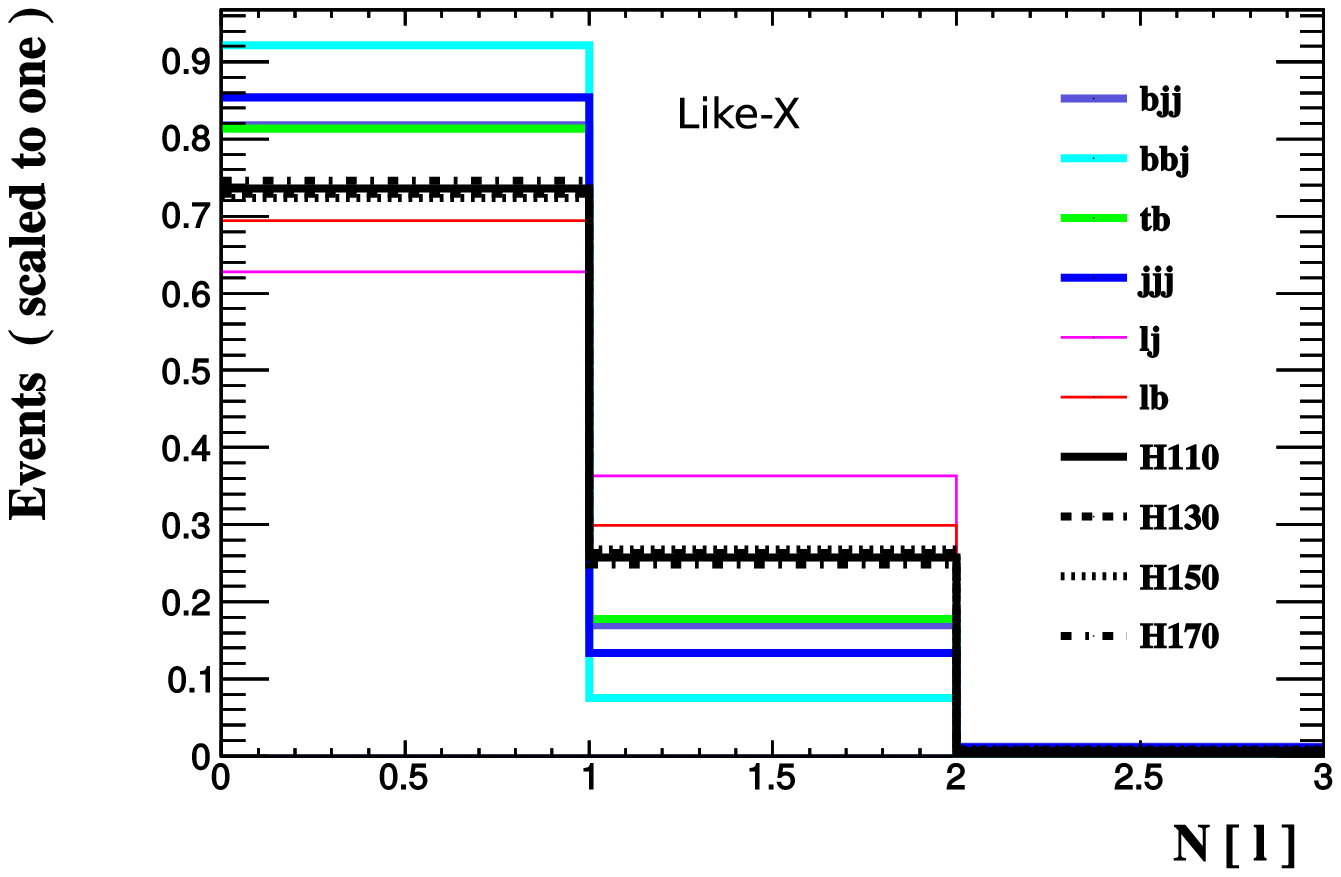}
	\includegraphics[scale=.45]{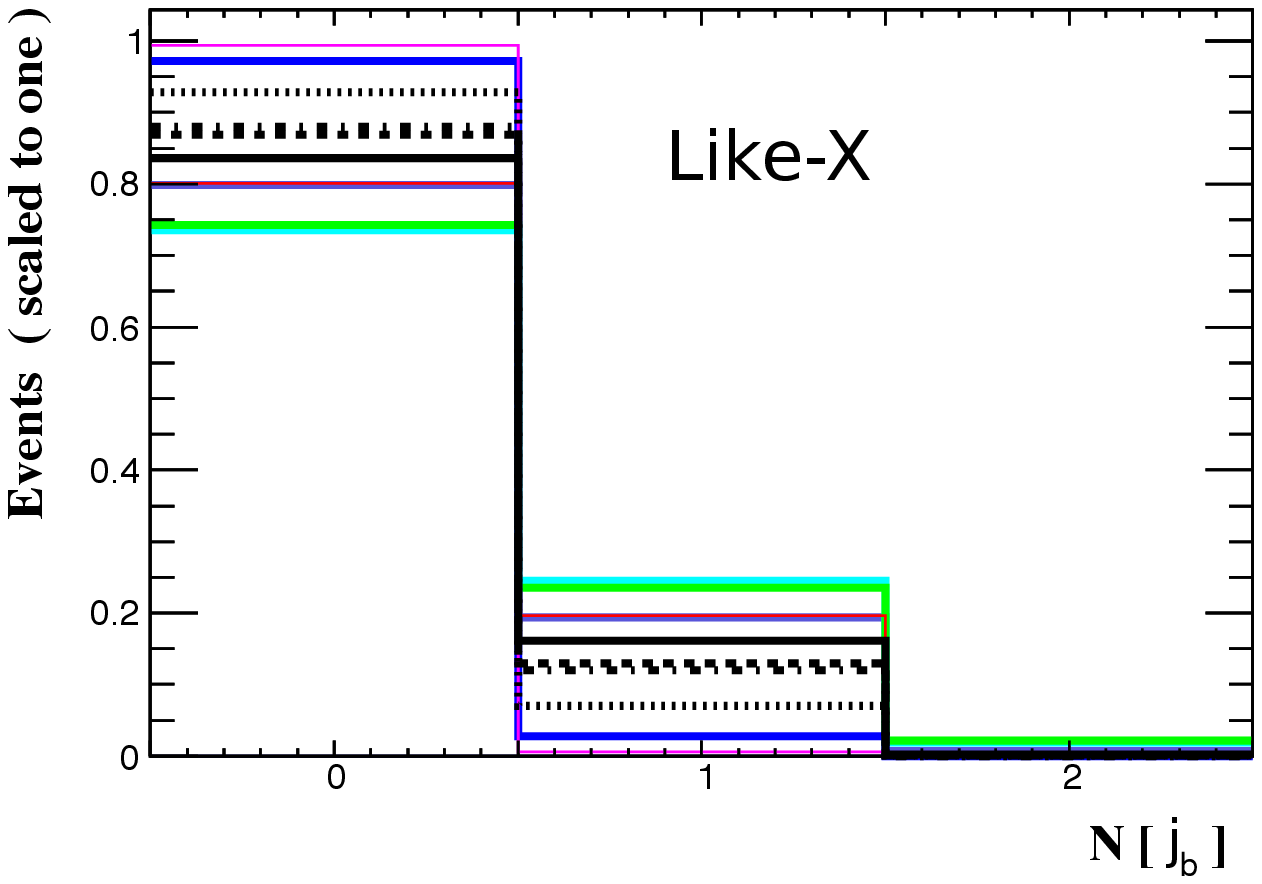}
        \caption{Distributions for the  process  $e^-q\to \nu_e H^- b$  followed by {$H^-\to \tau \bar{\nu_\tau}$}:
in the  left(right)  panel we present the number of leptons($b$-jets) per event. The like-X case is illustrated. The normalisation is to unity.
		}
        \label{Ntau1}
\end{figure}

\begin{figure}[!h]
        \centering
        \includegraphics[scale=.45]{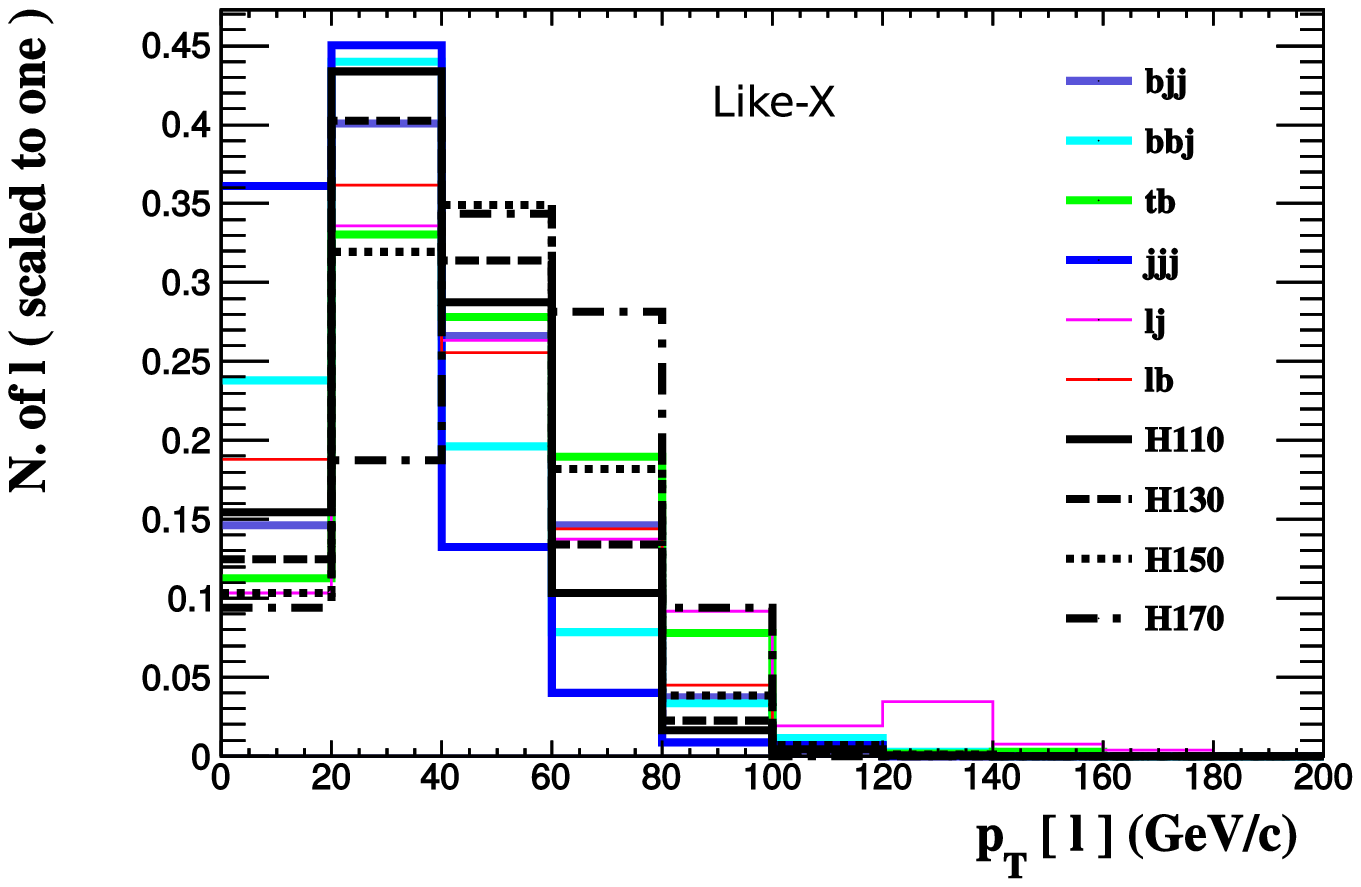}
        \includegraphics[scale=.45]{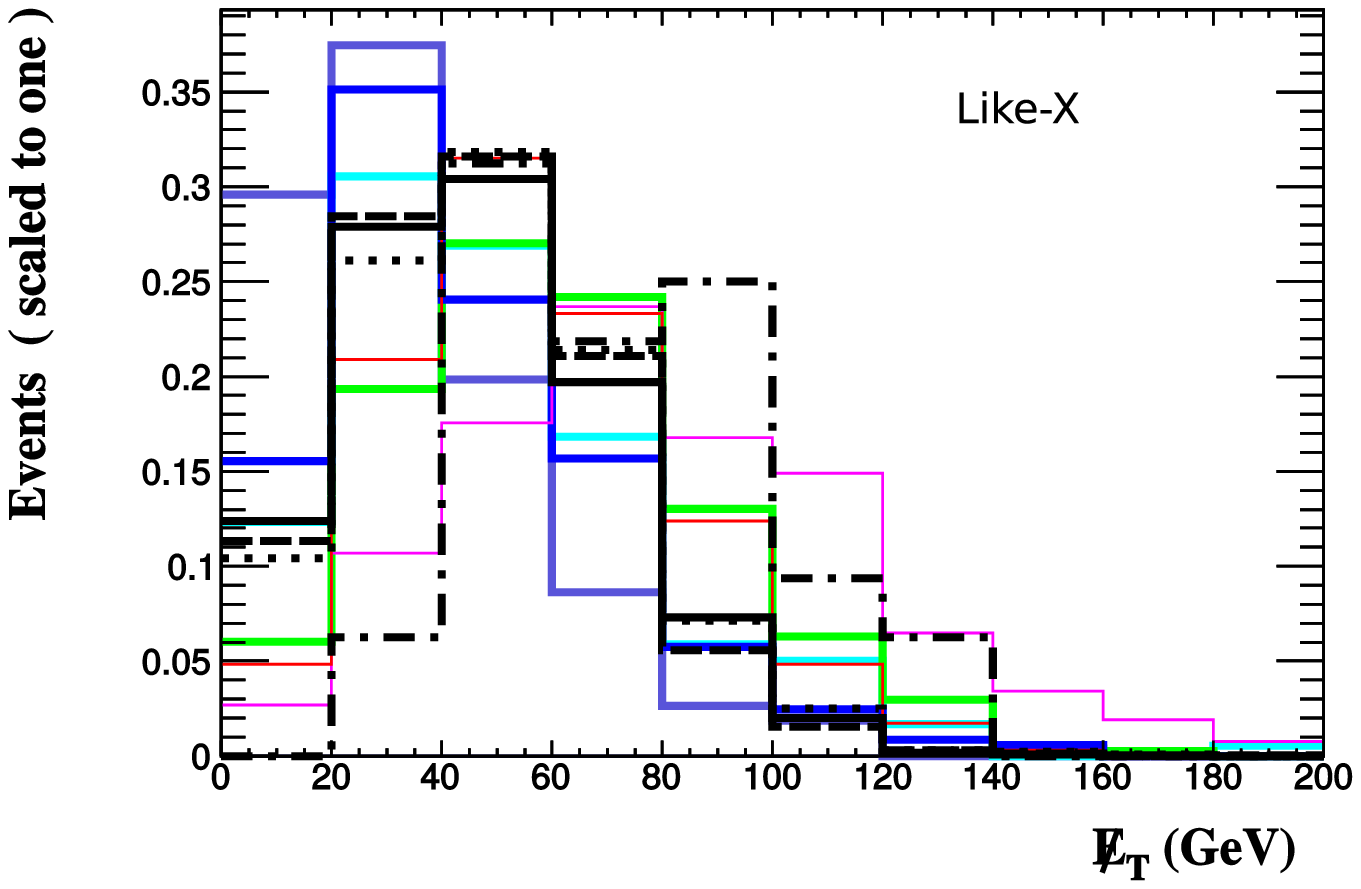}
        \caption{Distributions for the  process  $e^-q\to \nu_e H^- b$  followed by {$H^-\to \tau \bar{\nu}_\tau$}:
in the  left panel we present the 
transverse momentum of the lepton  while in the right  panel we present the total missing transverse energy. The like-X case is illustrated. The normalisation is to unity.
		}
        \label{PTtau2}
\end{figure}

\begin{figure}[!h]
        \centering
        \includegraphics[scale=.45]{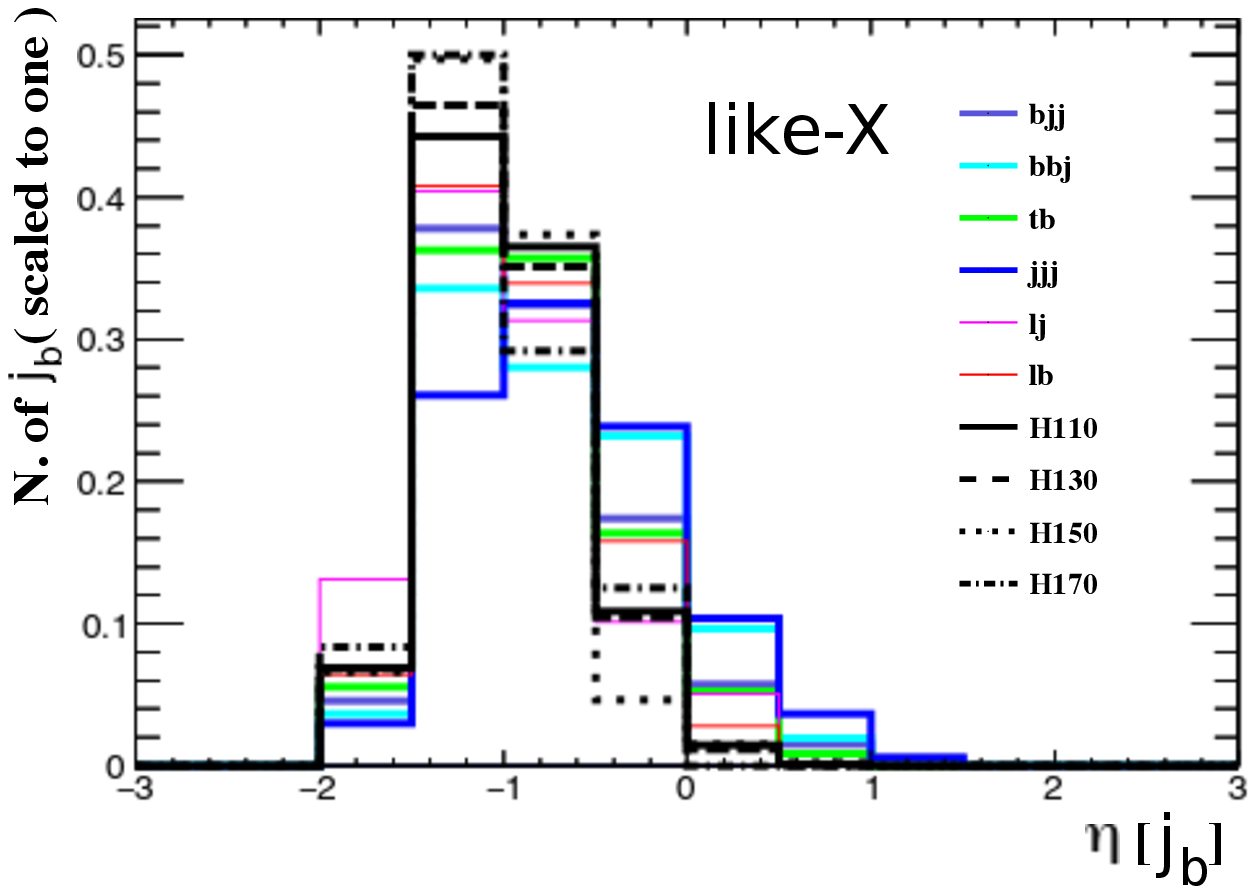}
	\includegraphics[scale=.45]{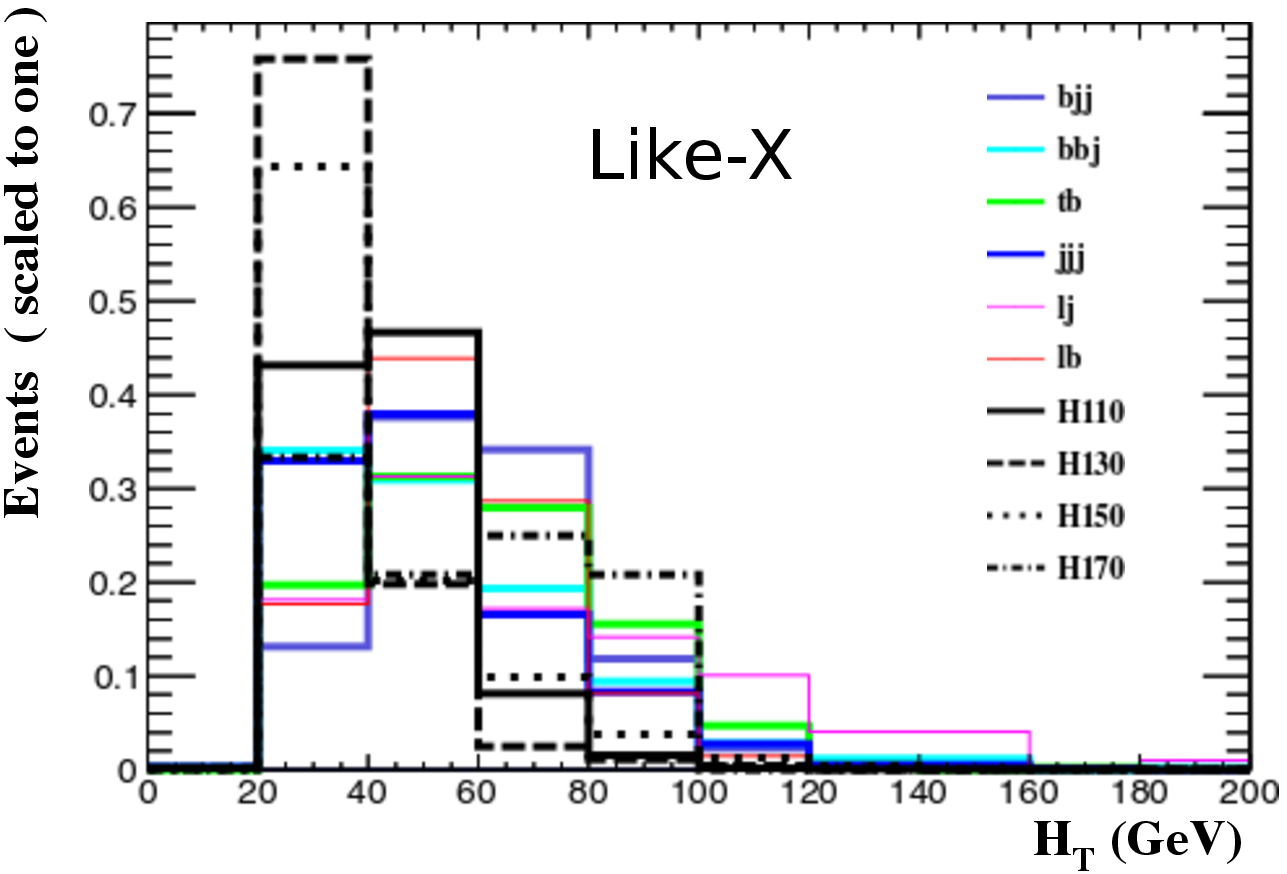}
        \caption{Distributions for the  process  $e^-q\to \nu_e H^- b$  followed by {$H^-\to \tau \bar{\nu}_\tau$}:
in the  left panel we present the pseudorapidity of the $b$ jet while in the right panel we present the total hadronic transverse  energy. The like-X case is illustrated. The normalisation is to unity.
           }
        \label{Ntau3}
\end{figure}

\begin{figure}[!h]
        \centering
        \includegraphics[scale=.45]{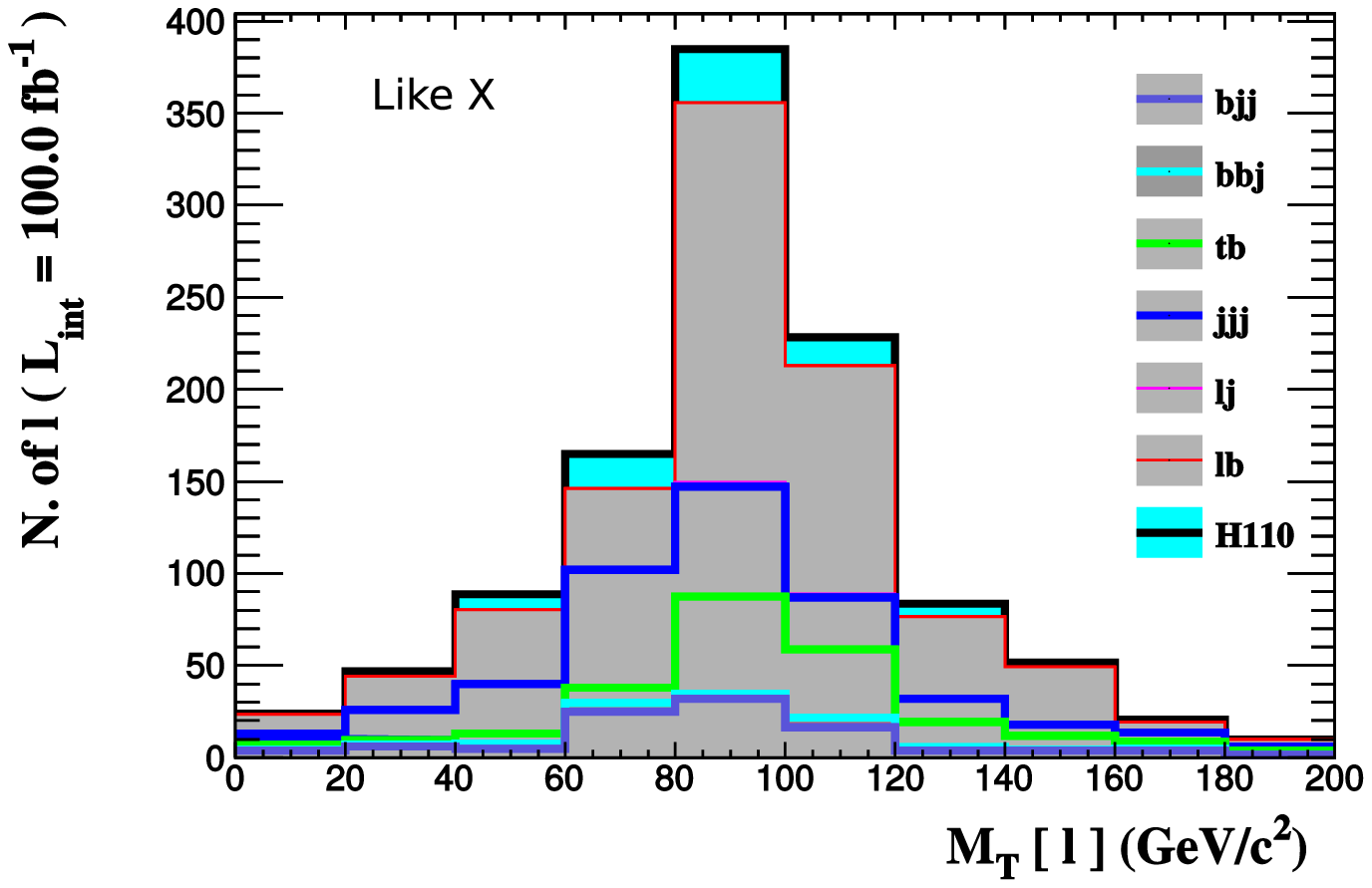}
	\includegraphics[scale=.45]{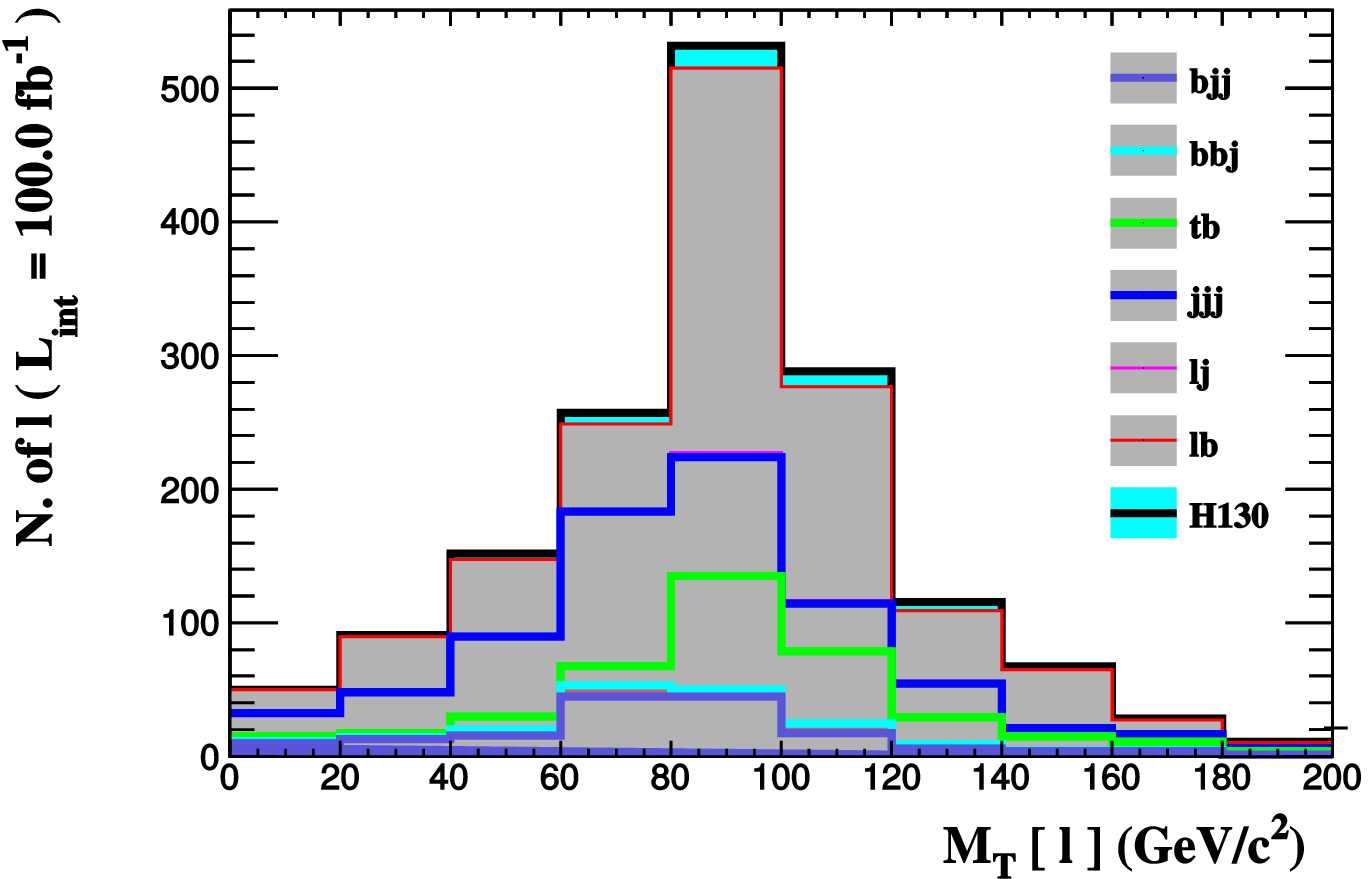}\\	
        \caption{Distributions for the  process  $e^-q\to \nu_e H^- b$  followed by {$H^-\to \tau\bar \nu_\tau$} in the transverse mass of the final state for $m_{H^\pm}=110$ GeV (left) and  $m_{H^\pm}=130$ GeV (right). The like-X case is illustrated. The normalisation is to the total event rate for $L=100$ fb$^{-1}$.
		}
        \label{Ntau4}
\end{figure}

\begin{enumerate}[I)]
	\item This first set of cuts is focused on selecting events with  one $b$-tagged jet and one lepton, by imposing $|\eta(b_{\rm tag},l)|<2.5$,
$P_T(b_{\rm tag},l)>20$  GeV and the isolation condition $\Delta R (b_{\rm tag};l)>0.5$ (see  Fig. \ref{Ntau1} for the histograms of the lepton and jet multiplicities.) Following this, we find that our signal has an efficiency of $14\%$ whereas the backgrounds $\nu_e\nu_l l j$ and $\nu_e\nu_l l b$ have rates of $23\%$ and $18\%$, respectively. The remaining noise shows an efficiency below $5\%$.    

	\item   The next set of cuts enables us to select a stiffer lepton and impose conditions on the  missing transverse energy which are adapted to the trial $H^\pm$ mass. 
We select events with  $P_T(l)>25(40)$ GeV and $E_T\hspace{-4mm}/>30(40)$ GeV for  $m_{H^\pm}=110$, $130$($150, 170$) GeV. Our signal presents an  efficiency of $70\%$ while  $80\%$ is the rate for  $\nu_e\nu_l l j$, $\nu_e\nu_l l b$ and $\nu_e t b$.  The remaining  backgrounds show efficiencies of $60\%$ or below  (see Fig. \ref{PTtau2}).

	\item Then, based on the left frame of   Fig.~\ref{Ntau3}, we require $|\eta(b_{\rm tag})|>0.5$. Furthermore, upon defining
the  total  hadronic transverse energy $H_T=\sum_{\rm hadronic}|P_T|$ in the final state,  based on the right frame of Fig.~\ref{Ntau3}, we select $H_T<60$ GeV. For our signal, these cuts are little  discriminatory and show an  efficiency of $75\%$.  However, for all  backgrounds, the efficiency is in general below  $50\%$. 

	\item Finally, we enforce the last selection by exploiting the transverse mass 
$M_T(l)^2=2 p_T(l) E_T\hspace{-4mm}/~~(1-\cos\phi)$, where $\phi$ is the relative azimuthal angle between $p_T(l)$ and $ E_T\hspace{-4mm}/$~~, a quantity which allows one to label  the candidate events reconstructing the charged Higgs boson mass. However, the existence of one additional neutrino in the final state ($\nu_e$) emerging from the primary hard  collision, alongside the two stemming from the    $\tau$ decay ($\nu_\tau$ and $ \nu_l$), generates a widening of the transverse mass distribution of the signal. Therefore, we make the following selection: $ m_{H^\pm} -50$ GeV $ <M_T(l)  <m_ {H^\pm} + 10 $ GeV (see Fig. \ref{Ntau4}). For this cut, our signal has a cumulative efficiency of $ 1 \% $, quite comparable to the efficiency of $ \nu l b $, which is  $ 0.9\% $. The rest of the backgrounds are instead below $0.2\%$ 
\end{enumerate}

The effectiveness of this selection strategy is confirmed by the final results in Tab. IV, wherein we present the signal and background rates along with the corresponding significances after Cuts I--IV for the usual values of luminosity. Again, also in the like-X case, good sensitivity  exists up to $H^\pm$ masses of 130 GeV. 
\section{Conclusions}

\begin{table} \label{taunu}
	\begin{center}
		\begin{tabular}{|r|c|c|r|r|r|r|c|}
		\hline
		\hline
		Signal & Scenario & Events (raw) & Cut I & Cut II & Cut III &Cut IV & ${\cal(S/\sqrt B)}
		_{100\ {\rm fb}^{-1}(1000\ {\rm fb}^{-1})[3000\ {\rm fb}^{-1}]} $ \\
		\hline
		\hline
		$\nu_e H^- q$ &X-110 &6480 & 178 & 124 &94 &67 &2.41 (7.61) [13.19] \\ 
			&X-130 &3390 & 75 & 54& 52 & 35 & 1.13	 (3.58) [6.2]\\
			&X-150 &880 &6 & 3& 2 & 2& 0.09 (0.29) [0.5]\\
			&X-170 & 20 & 0.4 & 0.3 & 0.2 & 0.09 & 0.01 (0.02) [0.04]\\
		\hline
		$\nu_e b b j$ &      &20170 &  85& 56 & 23 &13 &  \\ 
		\cline{1-7}
		$\nu_e bjj $ &       & 117559&   623& 340 & 122 & 84 & \\ 
		\cline{1-7}
		$\nu_e tb $ &        &48845 &  460 & 374 & 149 & 105& ${\cal B}=763$  \\ 
		\cline{1-7}
		$\nu_e jjj$ &     & 867000 &  981 & 596 & 267 & 162 & $\sqrt{\cal B}=27.62$ \\ 
		\cline{1-7}
		$\nu_e l \nu_l j$ &    &23700   & 29 & 26&  8& 5&  \\ 
		\cline{1-7}
		$\nu_e l \nu_l b$ &    & 40400 & 1500 & 1203 & 569& 392&  \\ 
		\hline
		\end{tabular}
		\caption{Significances obtained  after the sequential cuts described in the text for the signal process  $e^-q\to \nu_e H^- b$  followed by {$H^-\to \tau\bar \nu_\tau$}  for four BPs in the  2HDM-III  like-X. The simulation is done at detector level. {In the column Scenario,  the label X-110(130)[150]\{170\} means  $m_{H^\pm} =110$(130)[150]\{170\} GeV in the 2HDM-III like -X.} }
	\end{center}
\end{table}
In conclusion, we have assessed the potential of a possible future LHeC, obtained from crossing $e^-$ and $p$ beams in the  CERN tunnel currently hosting the LHC and previously LEP.
The foreseen beam energies are 60 GeV and 7 TeV, respectively. Such an environment is rather clean and, since it primarily relies on a charged $W^-$ current for the hard scattering, conducive to the production of a negatively charged Higgs boson, $H^-$.  This state is typical of 2HDMs and it is notoriously elusive at the LHC \cite{Akeroyd:2016ymd, Arhrib:2018ewj}, so that it is natural to assess the scope for its detection at the LHeC. As 2HDM theoretical framework we have adopted a 2HDM-III supplemented by a four-zero-texture in the Yukawa sector which enables one, firstly, to avoid imposing a $Z_2$ symmetry to prevent FCNCs and, secondly, to re-create the standard 2HDM setups, known as Type I,  II,  X and  Y, through suitable choices of the texture matrix elements. Such a scenario can realistically only afford one with LHeC sensitivity to rather light $H^\pm$ masses, i.e., well below the top mass. In this mass regime, though, we have established that 
the LHeC can access $H^\pm$ masses up to 130 GeV or so, for luminosity conditions already foreseen for such a machine. This assessment is essentially similar for all 2HDM-III incarnations, although sensitivity is primarily established in the like-I, -II and -Y cases via $H^-\to b\bar c$ and in the like-X case via $H^-\to \tau\bar\nu_\tau$ (assuming electron/muon decays of the $\tau$). The LHeC production mode is
     $e^-q\to \nu_e H^- q$, with $q=b$ being the dominant sub-channel, the latter being also induced by neutral Higgs boson exchange in $t$-channel (see Fig. \ref{Feynsignal}). 
Hence, on the one hand, one can exploit the very efficient $b$-tagging expected at the LHeC detectors in order to establish the two signals above and beyond a variety of background channels, which we have done here, while, on the other hand, one could attempt extracting the $\phi_i^0W^+H^-$ ($\phi_i^0=h,H,A$) vertex `directly' in LHeC production (unlike the LHC, where it can only be done `indirectly' in $H^-$ decays), which is what we shall do in a future publication.

\section*{Acknowledgements}
SM is financed in part through the NExT Institute.
 SM also acknowledges support from the UK STFC Consolidated grant ST/L000296/1 and
 the H2020-MSCA-RISE-2014 grant no.  645722 (NonMinimalHiggs). SR-N thanks the Southampton High Energy Physics Group for hospitality while parts of this work were completed. JH-S and CH have been supported by SNI-CONACYT
(M\'exico), VIEP-BUAP and  PRODEP-SEP (M\'exico)
under the grant `Red Tem\'atica: F\'{\i}sica del Higgs y del
Sabor'. SRN acknowledges a scholarship from
CONACYT (M\'exico). We all acknowledge useful discussions with Siba Prasad Das.

\end{document}